%% file: main.tex
\documentclass[10pt,twocolumn,letterpaper]{article}

\usepackage[pagenumbers]{cvpr} 
\usepackage{algorithm}
\usepackage{algpseudocode}

\input{preamble}
\definecolor{cvprblue}{rgb}{0.21,0.49,0.74}
\usepackage[pagebackref,breaklinks,colorlinks,allcolors=cvprblue]{hyperref}

\def\confName{CVPR}
\def\confYear{2027}

\title{MSRGF: Multi-Source Reconstruction with Reference-Guided Fusion for Incomplete Multimodal Sentiment Analysis}

\author{
Yan Zhuang$^{1,2}$ \quad
Minhao Liu$^{1}$ \quad
Yanru Zhang$^{1}$ \quad
Li Ma$^{1}$\\
Weilong Chen$^{1}$ \quad
Xingyu Wu$^{2}$ \quad
Jiawen Deng$^{1}$\thanks{Corresponding authors.} \quad
Fuji Ren$^{1}$\footnotemark[1]\\
$^{1}$University of Electronic Science and Technology of China\\
$^{2}$The Hong Kong Polytechnic University\\
{\tt\small
\{202211081370,202421080935\}@std.uestc.edu.cn, xingy.wu@polyu.edu.hk}\\
{\tt\small
\{minhaoliu,yanruzhang,weilongchen,dengjw,renfuji\}@uestc.edu.cn}
}

\begin{document}
\maketitle
\input{0_abstract}

\input{1_intro}
\input{2_related_works}
\input{3_methodology}

\input{4_experiments}
\input{5_conclusion}
{
    \small
    \bibliographystyle{ieeenat_fullname}
    \bibliography{main}
}

\input{X_suppl}

\end{document}

%% file: 0_abstract.tex
\begin{abstract}
Incomplete multimodal sentiment analysis (MSA) predicts sentiment from textual, acoustic, and visual features when some observations are missing. Existing reconstruction methods estimate missing features from available modalities and combine the resulting information for sentiment prediction. However, different available modalities can produce distinct estimates of the same modality, while direct fusion often leaves the relationships among these estimates underused. We propose \textbf{MSRGF} (\textbf{M}ulti-\textbf{S}ource \textbf{R}econstruction with \textbf{R}eference-\textbf{G}uided \textbf{F}usion), which retains separate estimates and exploits their relationships during fusion. Specifically, MSRGF comprises three components: (1) \textbf{Source-Wise Reconstruction (SWR)}, which uses modality-specific queries and a shared generator to produce a separate estimate of each modality from every available modality; (2) \textbf{Reference-Guided Fusion (RGF)}, which uses relationships among estimates and available observations to construct a weighted reference, selectively refines it using their differences, and incorporates observed features when available; and (3) \textbf{Global Context Modulation (GCM)}, which uses global context to adjust modality summaries for sentiment prediction. Extensive experiments on three datasets show that MSRGF outperforms existing methods, achieving average accuracy improvements of 1.9\% on MOSEI under random within-modality missingness, 0.6\% under random whole-modality missingness and 1.3\% under fixed whole-modality missingness.
\end{abstract}

%% file: 1_intro.tex
\section{Introduction}
\label{sec:introduction}

Multimodal sentiment analysis (MSA) combines textual, acoustic, and visual information to predict sentiment \cite{liang2023multizoo,liang2024foundations,zadeh2016mosi}. It supports applications such as human--computer interaction and video understanding \cite{Lian2026merbench}. In practice, transcripts may be incomplete \cite{liu2021face,jaiswal2020privacy}, speech segments may be unavailable, and visual observations may be interrupted by occlusion or tracking failures \cite{zhuang2026iie}. Such disruptions can cause missing features within a modality or an entirely missing modality, making sentiment prediction more challenging \cite{zhao2021mmin,guo2024mplmm,zhuang2026tmdc,wei2023mmanet}.

\begin{figure}[t]
    \centering
    \begin{minipage}[t]{0.47\textwidth}
        \centering
        \includegraphics[width=\linewidth]{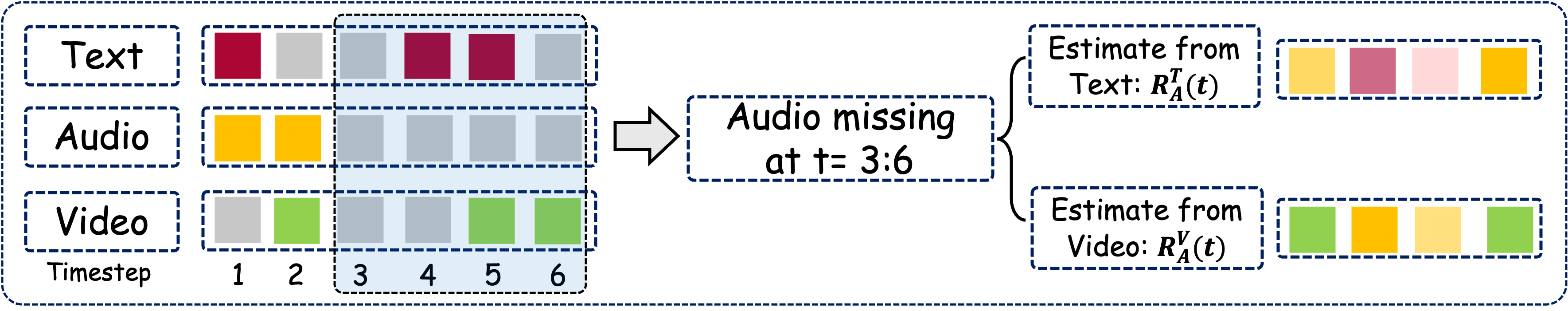}
        {\footnotesize\bfseries
        (a) Incomplete Inputs and Source-wise Estimates}
    \end{minipage}
    \par\vspace{2mm}
    \begin{minipage}[t]{0.215\textwidth}
        \centering
        \includegraphics[width=\linewidth]{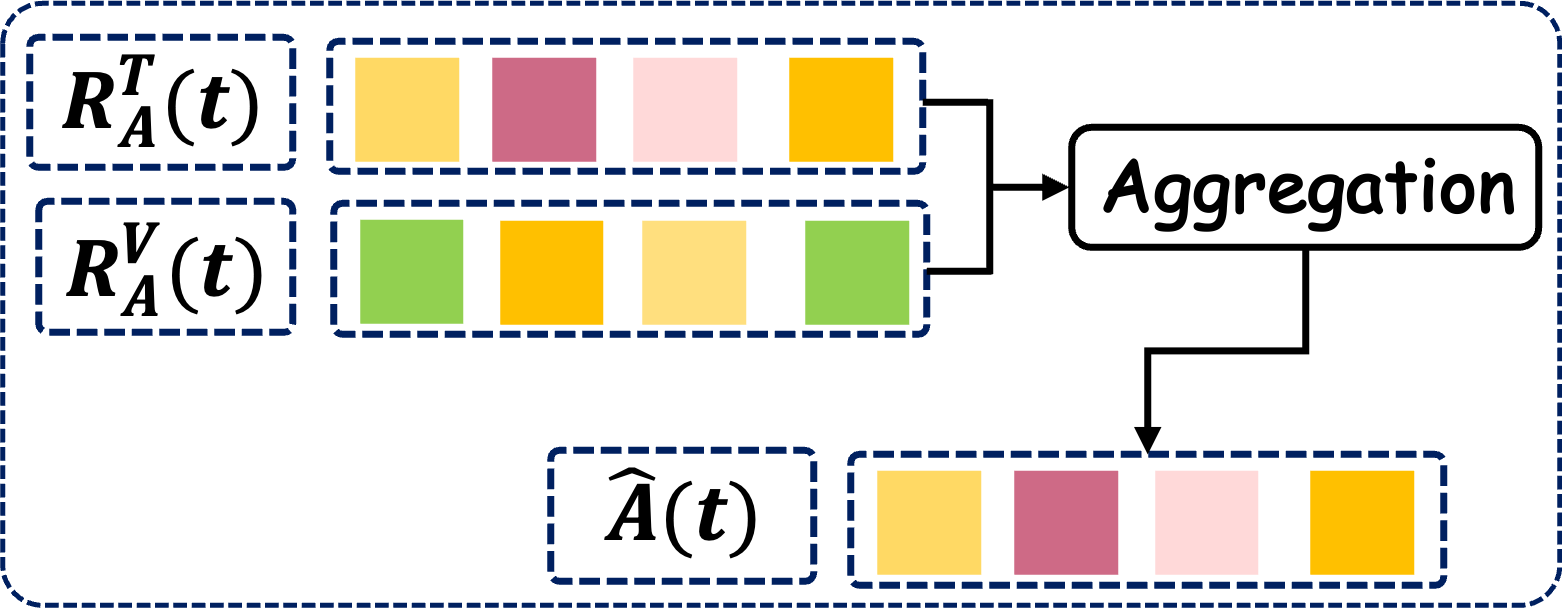}
        {\footnotesize\bfseries
        (b) Conventional Fusion}
    \end{minipage}\hfill
    \begin{minipage}[t]{0.255\textwidth}
        \centering
        \includegraphics[width=\linewidth]{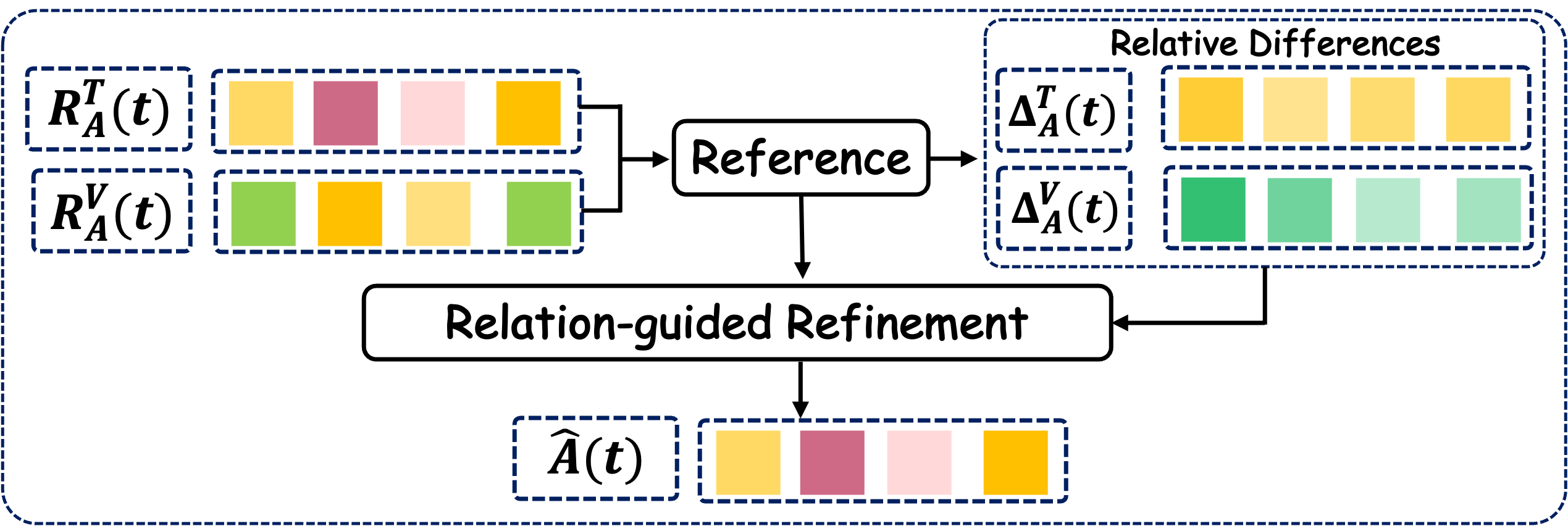}
        {\footnotesize\bfseries
        (c) Our MSRGF}
    \end{minipage}
    \caption{Motivation of MSRGF. (a) Under incomplete multimodal observations, available textual and visual information can yield distinct estimates of the same missing acoustic features. (b) Direct aggregation combines these estimates into a single representation without explicitly examining their relationships. (c) MSRGF retains source-wise estimates, constructs a weighted reference, and selectively incorporates transformed deviations from the reference to refine the target representation.}
    \label{fig:motivation}
\end{figure}

Reconstruction methods address missing observations by estimating unavailable representations from the remaining modalities \cite{lian2023gcnet,wang2023imder,wang2023dicmor,pham2019found,wang2023multi} and integrating them with observed features for sentiment prediction. Prior work has explored cross-modal translation \cite{pham2019mctn,liu2024modality}, representation reconstruction \cite{zhao2021mmin,yuan2021transformer}, distribution modeling \cite{wang2023dicmor}, diffusion models \cite{wang2023imder}, and generative prompts \cite{guo2024mplmm,wang2026api}. When multiple modalities are available, each can provide a different estimate of the same modality. These estimates can be combined to construct a recovered representation, but their individual relationships are often not explicitly used to guide the fusion process.

As illustrated in Figure~\ref{fig:motivation}, different modalities may have different missing patterns over time. When acoustic information is partially missing, the available textual and visual observations can provide separate estimates of the missing features. These estimates may agree in some features while differing in others. Simply aggregating them into a single representation may leave their relative differences insufficiently exploited. This motivates retaining the separate estimates and learning how their relationships contribute to the final target representation.

To this end, we propose \textbf{MSRGF}, a Multi-Source Reconstruction with Reference-Guided Fusion framework for incomplete MSA. MSRGF consists of three modules. First, \emph{Source-Wise Reconstruction} (SWR) employs modality-specific queries and a shared cross-attention generator to produce a separate estimate of each modality from every available modality, retaining these estimates for subsequent fusion. Building on these estimates, \emph{Reference-Guided Fusion} (RGF) compares them with one another and with available information to construct a weighted reference. It then uses source-wise weighting and feature-wise gating to selectively integrate their transformed differences, while incorporating observed features of the estimated modality when available. Finally, \emph{Global Context Modulation} (GCM) summarizes the enhanced modality representations and uses global context to adjust them for sentiment prediction. Together, these modules enable MSRGF to exploit multi-source reconstruction information under diverse missingness conditions.

Our main contributions are summarized as follows:
\begin{itemize}
    \item We propose MSRGF for incomplete MSA, which generates separate estimates from available modalities and exploits their relationships to enhance modality representations.

    \item We introduce RGF to construct a weighted reference using relationships among source-wise estimates, and selectively incorporate their transformed differences and observed features to enhance modality representations.

    \item Extensive experiments on three benchmark datasets under fixed whole-modality combinations, random whole-modality missingness, and random within-modality missingness demonstrates the effectiveness of MSRGF.
\end{itemize}

%% file: 2_related_works.tex
\section{Related Work}
\label{sec:civic_related}

\subsection{Incomplete MSA}
Incomplete MSA aims to maintain sentiment prediction when textual, acoustic, or visual information is partially or entirely unavailable \cite{xu2024leveraging, xu2025distilled, lin2026cyin, xu2026mcmoe,chen2026sample,meng2026general}. Existing methods address this challenge through knowledge distillation \cite{pan2024disentangled, li2024corrkd, zhuang2025cmad}, representation enhancement \cite{zhuang2026hyper}, and missing-feature reconstruction \cite{guo2024mplmm}. Knowledge-distillation methods \cite{li2024umdf, wei2023mmanet, li2024corrkd, wang2020multimodal, stroud2020d3d, garcia2018modality, yang2023knowledge, lao2023unikd} leverage a teacher model trained on complete information to transfer information to the student model trained on incomplete inputs.  For example, CorrKD \cite{li2024corrkd} introduces sample- and category-level distillation, UMDF \cite{li2024umdf} combines self-distillation with cross-modal interaction, and CMAD \cite{zhuang2025cmad} transfers feature correlations through modality-aware optimization. Other methods improve incomplete-input representations through hierarchical learning in HRLF \cite{li2024hrlf}, language-guided correction in LNLN \cite{zhang2024lnln}, and cross-sample enhancement in HME \cite{zhuang2026hyper} and IIE \cite{zhuang2026iie}.

Reconstruction-based methods recover missing modality representations from available observations. MCTN \cite{pham2019mctn} learns cross-modal representations through cyclic translation, while MMIN \cite{zhao2021mmin} reconstructs missing representations using available modalities. DiCMoR \cite{wang2023dicmor} models cross-modal distribution transfer, IMDer \cite{wang2023imder} employs conditional diffusion, and MPLMM \cite{guo2024mplmm} uses generative prompts for feature recovery. More recent methods, including T$^2$DR \cite{lin2025t2dr} and API \cite{wang2026api}, further explore feature prediction and prototype adaptation under incomplete inputs. These methods provide different ways to estimate missing representations from available information. MSRGF further considers the relationships among separate estimates of the same modality, retaining them during reconstruction and using them to construct the enhanced representation.

\subsection{Multimodal Interaction and Fusion}

Multimodal interaction and fusion aim to combine information from different modalities \cite{YangZQYLC26,DengGZPXKWL26}. MISA \cite{hazarika2020misa} learns modality-invariant and modality-specific representations, while ALMT \cite{zhang2023almt} introduces language-guided attention for cross-modal interaction. CACR \cite{wang2026cacr} incorporates conflict-aware cross-reconstruction and sentiment refinement, and EBMC \cite{he2026ebmc} combines modality enhancement with balanced optimization to improve multimodal learning. For incomplete inputs, fusion methods further consider the availability and quality of modality representations. RECAP \cite{huang2026recap} integrates affective recovery with mutual-information-guided attention, while DEAR \cite{zou2026dear} uses reconstruction fidelity and uncertainty to regulate recovered features. ComP \cite{he2026comp} employs cross-modal prompts and knowledge propagation to enhance representations. These methods improve multimodal fusion by considering cross-modal interactions, reconstruction quality, or uncertainty. MSRGF focuses on the relationships among separate estimates of the same modality produced by different available modalities. RGF uses these relationships to construct a weighted reference and selectively refine it through the differences among estimates.

\begin{figure*}[t]
    \centering
    \includegraphics[width=0.95\linewidth]{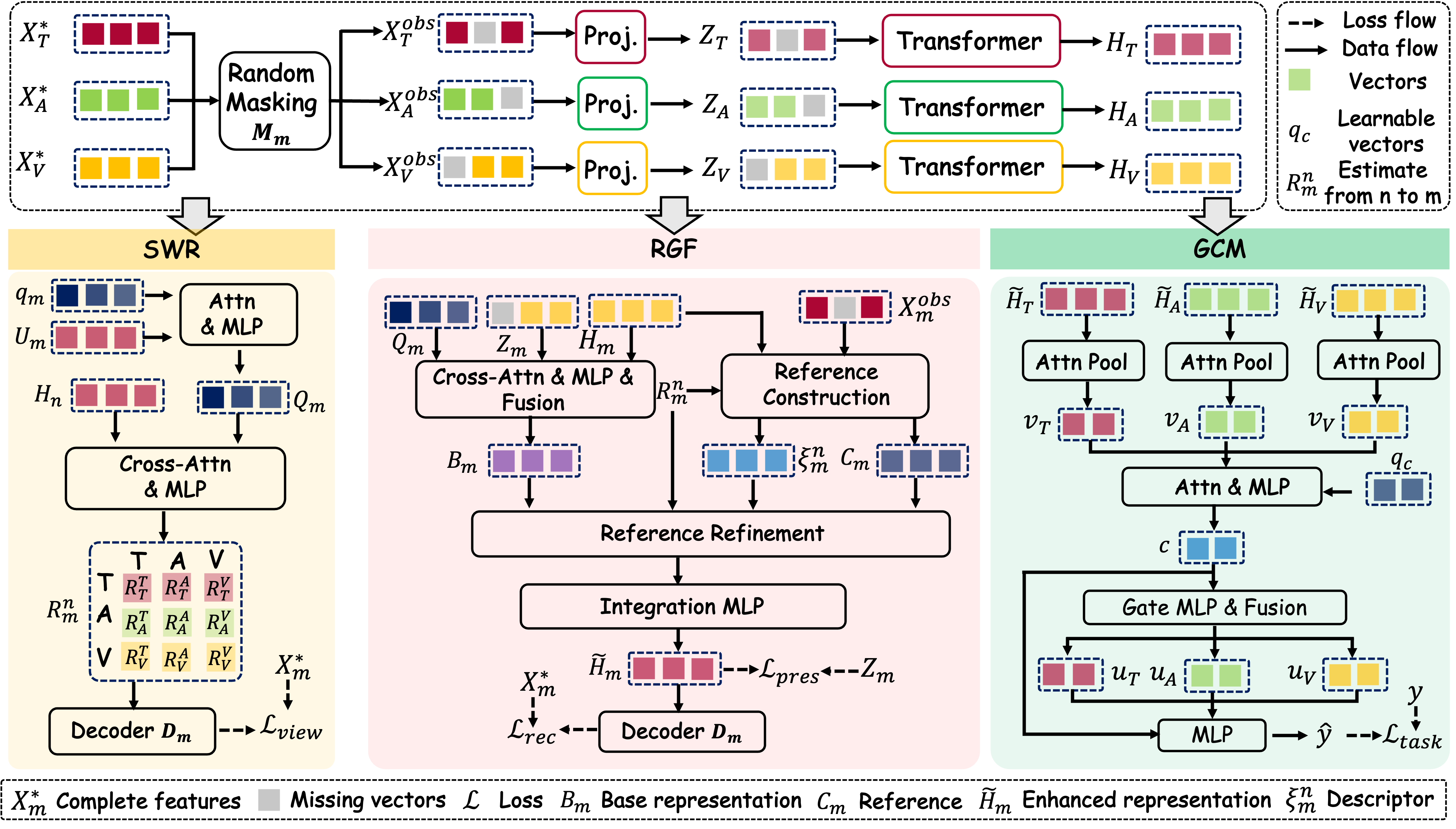}
    \caption{Overview of MSRGF. SWR produces a separate estimate of each modality from every available modality. RGF constructs a weighted reference, selectively refines it using differences among estimates, and integrates the result with a base representation that incorporates observed features when available. GCM uses global context to adjust modality summaries for sentiment prediction.}
    \label{fig:framework}
\end{figure*}

%% file: 3_methodology.tex
\section{Methodology}
\label{sec:msrgf_method}

\subsection{Problem Formulation}
\label{sec:msrgf_overview}
We consider incomplete MSA with textual, acoustic, and visual modalities, denoted by $\mathcal M=\{T,A,V\}$. For each modality $m\in\mathcal M$, $X_m^{\mathrm{obs}}\in\mathbb R^{L\times d_m}$ denotes its observed feature sequence, where $L$ is the aligned sequence length and $d_m$ is its feature dimension. The observation mask $M_m\in\{0,1\}^L$ identifies observed positions, and the validity mask $P_m\in\{0,1\}^L$ excludes padding, with $M_m\leq P_m$. Features at missing positions are set to zero; $M_m=\mathbf 0$ indicates that modality $m$ is entirely missing. Each sample retains at least one observed position, so the set of available modalities is nonempty: $\mathcal O=\{n\in\mathcal M:\|M_n\|_1>0\}$. The objective is to predict sentiment $\hat y$ from $\{X_m^{\mathrm{obs}},M_m,P_m\}_{m\in\mathcal M}$. In controlled missingness experiments, complete features $X_m^\star$ are retained only as reconstruction targets during training.

\subsection{Framework Overview}
\label{sec:msrgf_framework}
Different available modalities can provide different estimates of the same missing information. To use the relationships among these estimates, MSRGF keeps them separate until fusion (Figure~\ref{fig:framework}). \emph{Source-Wise Reconstruction} (SWR) first produces an estimate of each modality from every available modality. \emph{Reference-Guided Fusion} (RGF) then compares the estimates to determine their contributions, forms a weighted reference, and uses their remaining differences to refine the representation. Observed features of the estimated modality, when present, also participate in this fusion. Finally, \emph{Global Context Modulation} (GCM) combines the three enhanced representations for sentiment prediction.

\subsection{Source-Wise Reconstruction}
\label{sec:msrgf_swr}
SWR needs to answer two questions for each estimate: what information is available, and which modality and temporal positions should be estimated. We use $n\in\mathcal O$ for an available modality providing information and $m\in\mathcal M$ for the modality being estimated.

Each modality first maps its observed features to a common hidden dimension $d$ through a linear modality-specific projection followed by LayerNorm, denoted by $\phi_m$. A temporal Transformer $E_m$ then reads the observed positions to produce their contextual representations:
\begin{equation}
\label{eq:msrgf_encoding}
\begin{aligned}
Z_m&=M_m\odot\phi_m(X_m^{\mathrm{obs}}),\\
H_m&=M_m\odot E_m(Z_m+e^{\mathrm{pos}};M_m).
\end{aligned}
\end{equation}
Here $e^{\mathrm{pos}}$ denotes positional embeddings, and $M_m$ prevents missing positions from contributing attention keys or values. The encoded features $H_n$ will provide the information read from available modality $n$. We also retain $Z_m$ so that observed features of modality $m$ can enter RGF directly. If $M_m=\mathbf 0$, the temporal attention is skipped and $H_m$ is set to zero.

The features of modality $m$ cannot define the reconstruction positions when that modality is entirely missing. SWR therefore constructs a query $Q_m$ using modality identity $e_m^{\mathrm{mod}}$, position $e_t^{\mathrm{pos}}$, and observation state $e_{M_m(t)}^{\mathrm{mask}}$. At position $t$, these three embeddings form: $q_m^0(t)=e_t^{\mathrm{pos}}+e_m^{\mathrm{mod}}+e_{M_m(t)}^{\mathrm{mask}}$. A learned prompt bank $U_m$ contains additional trainable vectors for modality $m$. The initial query attends to this bank, and a shared MLP $f_Q$ produces:
\begin{equation}
\label{eq:msrgf_query_main}
Q_m(t)=P_m(t)f_Q\!\left(q_m^0(t)+
\operatorname{Attn}(q_m^0(t),U_m,U_m)\right).
\end{equation}
Thus, $Q_m$ specifies what SWR will estimate without reading missing feature values, while $H_n$ supplies the observations from which the estimate can be inferred.

\textbf{Multi-source estimates:} For each available modality $n$, the query $Q_m$ reads $H_n$ through a shared two-block cross-attention generator $\mathcal G$. Its output passes through a modality-specific MLP $f_m^{\mathrm{tar}}$:
\begin{equation}
\label{eq:msrgf_views}
R_m^n=P_m\odot f_m^{\mathrm{tar}}\!\left(
\mathcal G(Q_m+e_m^{\mathrm{mod}},H_n+e_n^{\mathrm{mod}};M_n)
\right).
\end{equation}
Here $Q_m$ supplies the queries, and the observed positions in $H_n$ supply the keys and values. The generator is shared across all modality pairs; for a fixed $m$, every available modality also uses the same output MLP $f_m^{\mathrm{tar}}$. Hence the estimates $R_m^n\in\mathbb R^{L\times d}$ have the same target positions and hidden dimension, while preserving which available modality produced each estimate. For each modality $m$, a modality-specific linear decoder $D_m:\mathbb R^d\to\mathbb R^{d_m}$ is shared across its estimates and final enhanced representation. It supports reconstruction supervision and the comparison
with observed features in RGF. If modality $m$ is partially observed, it can also contribute an estimate through $n=m$. SWR passes all $\{R_m^n\}_{n\in\mathcal O}$ to RGF.

\subsection{Reference-Guided Fusion}
\label{sec:msrgf_rgf}
SWR provides several estimates of modality $m$, but they need not contribute equally. Combining them immediately would discard the distinction between estimates, while treating them equally would ignore differences. RGF first uses their relationships to form a reference, then uses their deviations from that reference to construct a refinement. When modality $m$ has observed features, these also help determine how the estimates are used.

For a partially observed modality, its own observed features provide information that need not be inferred from other modalities. RGF therefore builds a base representation $B_m$ with a direct path from $Z_m$. To supply features at positions without observations, a separate cross-attention block $\mathcal C_{\mathrm{anc}}$ reads $H_m$ using $Q_m$, followed by a shared
MLP $f_{\mathrm{anc}}$:
\begin{equation}
\label{eq:msrgf_anchor_aux}
a_m=f_{\mathrm{anc}}\!\left(\mathcal C_{\mathrm{anc}}(Q_m,H_m;M_m)\right).
\end{equation}
The base representation combines this result with the available observed features:
\begin{equation}
\label{eq:msrgf_anchor}
B_m=P_m\odot\left[M_m\odot(Z_m+\eta_a a_m)+(1-M_m)\odot a_m\right]
\end{equation}
Here $\eta_a$ is a fixed scaling coefficient. At observed positions, $Z_m$ enters directly and $a_m$ provides a scaled update; elsewhere, $B_m$ uses $a_m$. When modality $m$ is entirely missing, the cross-attention update is skipped, but the query residual and feed-forward operations still produce $a_m$. As a result, $B_m$ can guide subsequent weighting even without observed features of modality $m$.

\textbf{Reference construction:} The weight assigned to $R_m^n$ should account for more than its value at a single position. We therefore compare the estimate with the modality that produced it, with other estimates of $m$, and, when possible, with observed features of $m$. These comparisons form a relation descriptor $\xi_m^n$ for each estimate.

First, we pool $R_m^n$ over valid positions and $H_n$ over observed positions. After projecting the two pooled representations to the same dimension, let them be $a_m^n$ and $b_n$. A learned MLP $f_{\mathrm{src}}$ measures their interaction using the representations themselves, their product, and their absolute difference:
\begin{equation}
\label{eq:msrgf_source_relation}
\mathbf {c}_{m,n}^{\mathrm{src}}=f_{\mathrm{src}}\!\left(
[a_m^n,b_n,a_m^n\odot b_n,|a_m^n-b_n|]
\right).
\end{equation}
Second, we measure how $R_m^n$ differs from other estimates by averaging the cosine distances between their pooled representations. Third, when modality $m$ is observed and $n\ne m$, we compare $D_m(R_m^n)$ with $X_m^{\mathrm{obs}}$ at observed positions using mean cosine similarity. These two cosine measurements are used as comparison scores with stopped gradients; the source--estimate interaction remains trainable. When a comparison is unavailable, its score is set to zero and an indicator records that it was not performed.

The three comparisons, observation coverage of $m$ and $n$, and their validity indicators form the descriptor:
\begin{equation}
\label{eq:msrgf_descriptor}
\xi_m^n=\big[c_{m,n}^{\mathrm{tar}},\mathbf{c}_{m,n}^{\mathrm{src}},
 c_{m,n}^{\mathrm{div}},\rho_m,\rho_n,v_{m,n}^{\mathrm{tar}},
 v_n^{\mathrm{src}},v_n^{\mathrm{div}}\big].
\end{equation}
Here $c^{\mathrm{tar}}$ and $c^{\mathrm{div}}$ are the observed-feature agreement and cross-estimate difference scores, respectively, and $\rho_m=\|M_m\|_1/\max(\|P_m\|_1,1)$ denotes observation coverage. The three binary indicators $v_{m,n}^{\mathrm{tar}}$,$v_n^{\mathrm{src}}$, and $v_n^{\mathrm{div}}$ specify whether the corresponding comparisons are available, and the descriptor is calculated once for each estimate and reused across its temporal positions. The definitions and the exact comparison formulas are given in the Appendix.

RGF then combines this descriptor with the current estimate $R_m^n(t)$ and base representation $B_m(t)$ to calculate a weight for every available modality:
\begin{equation}
\label{eq:msrgf_alpha}
\alpha_m^n(t)=\operatorname{softmax}_{n\in\mathcal O}
 f_\alpha\!\left([R_m^n(t),B_m(t),\xi_m^n]+e_m^\alpha\right).
\end{equation}
Here $e_m^\alpha$ is a learned modality embedding that allows the shared scoring MLP to distinguish the modality being estimated. The normalized weights give the reference:
\begin{equation}
\label{eq:msrgf_reference}
C_m(t)=\sum_{n\in\mathcal O}\alpha_m^n(t)R_m^n(t).
\end{equation}
Thus, $C_m$ combines the estimates according to their relationships with the available information. The original $R_m^n$ remain available, so RGF can still examine how each estimate differs from this reference.

\textbf{Reference refinement:} The reference summarizes the estimates, but does not explicitly represent how any one of them deviates from it. For each available modality, RGF therefore computes $\Delta_m^n=R_m^n-C_m$ and transforms it with a shared nonlinear MLP $F_{\mathrm{diff}}$. Because these deviations may contain both useful information and reconstruction errors, the model learns how much each transformed deviation should contribute.

Two shared MLPs ($f_\beta$ and $f_G$) control this contribution. The source-wise weights $\beta_m^n(t)$ select among available modalities, while the feature-wise gates $G_m^n(t)$ select dimensions within each transformed deviation:
\begin{equation}
\label{eq:msrgf_diff_weights}
\begin{aligned}
\beta_m^n(t)&=\operatorname{softmax}_{n\in\mathcal O}f_\beta\!\left(
[R_m^n(t),B_m(t),\xi_m^n]+e_{m,n}^\beta\right),\\
G_m^n(t)&=\sigma\!\left(f_G\!\left(
[C_m(t),B_m(t),R_m^n(t),\xi_m^n]+e_{m,n}^G\right)\right).
\end{aligned}
\end{equation}
The learned embeddings $e_{m,n}^{\beta}$ and $e_{m,n}^{G}$ distinguish different modality pairs in the two shared MLPs. Unlike $\alpha_m^n$, which constructs the reference, $\beta_m^n$ controls the contribution of deviations. The gate $G_m^n\in\mathbb R^{L\times d}$ additionally adjusts individual feature dimensions using the reference itself. The refinement is then:
\begin{equation}
\label{eq:msrgf_difference}
N_m=\mathbb I[|\mathcal O|>1]\,P_m\odot
\sum_{n\in\mathcal O}\beta_m^n
\big[G_m^n\odot F_{\mathrm{diff}}(\Delta_m^n)\big].
\end{equation}
By transforming and weighting the deviations before aggregation, RGF can construct an update beyond the weighted reference. When only one modality is available, $C_m=R_m^n$ and the refinement is set to $N_m=0$.

\textbf{Integration:} RGF now has three representations with different roles: $B_m$ retains observed features where available, $C_m$ combines the separate estimates, and $N_m$ contains the selected difference-based update. The enhanced representation starts from $B_m$, moves toward $C_m$, and adds $N_m$ through learned gates:
\begin{equation}
\label{eq:msrgf_enhanced}
\widetilde H_m=P_m\odot\big[
B_m+g_m^{\mathrm{rec}}\odot(C_m-B_m)
+g_m^{\mathrm{enh}}\odot N_m\big].
\end{equation}
To obtain these gates, a shared MLP $f_{\mathrm{int}}$ receives $B_m,C_m,N_m,M_m$ and a relation summary $\bar\xi_m(t)=\sum_{n\in\mathcal O}\alpha_m^n(t)\xi_m^n$. It outputs two sigmoid gates $[r_m^{\mathrm{rec}},r_m^{\mathrm{enh}}]=\sigma(f_{\mathrm{int}}([B_m,C_m,N_m,\bar\xi_m,M_m]))$. The observation mask then adjusts them:
\begin{equation}
\label{eq:msrgf_gates}
\begin{aligned}
g_m^{\mathrm{rec}}&=(1-M_m)+\eta_gM_m\odot r_m^{\mathrm{rec}},\\
g_m^{\mathrm{enh}}&=(1-M_m+\eta_g M_m)\odot r_m^{\mathrm{enh}}.
\end{aligned}
\end{equation}
Here $\eta_g$ is a fixed scaling coefficient for updates at observed positions. At missing positions, $g_m^{\mathrm{rec}}=1$: the direct contribution of $B_m$ cancels, and the output becomes $C_m+g_m^{\mathrm{enh}}\odot N_m$. At observed positions, the path from $Z_m$ remains available through $B_m$, while both updates are scaled by $\eta_g$. In this way, RGF uses reconstruction where observations are missing and retains a direct observation path where they are present. For example, if audio is entirely missing but text and vision are available, the two estimates $R_A^T$ and $R_A^V$ first form $C_A$; their separate deviations then contribute to $N_A$, which refines the missing audio representation. The same computation retains a direct observation path for modalities that are only partially missing.

\subsection{Global Context Modulation}
\label{sec:msrgf_gcm}
RGF produces one enhanced sequence $\widetilde H_m$ for each modality. To obtain a sample-level sentiment prediction, GCM first summarizes each sequence over its valid positions:
\begin{equation}
\label{eq:msrgf_pool}
v_m=\operatorname{AttnPool}_m(\widetilde H_m;P_m).
\end{equation}
A learned global query then attends to the three summaries $\mathbf H_{\mathrm{sum}}=\operatorname{Stack}(v_T,v_A,v_V)$ through attention $\operatorname{MHA}_c$, producing $c=f_c\!\left(\operatorname{MHA}_c(q_c,\mathbf H_{\mathrm{sum}},\mathbf H_{\mathrm{sum}})\right)$. Here $f_c$ is the MLP, and this context $c$ provides information from all three enhanced modalities. GCM also adjusts each individual summary using the same context. It first projects $c$ to $s_m=W_m^{\mathrm{back}}c$ and uses the relationship between $v_m$ and $s_m$ to predict a feature-wise gate:
\begin{equation}
\label{eq:msrgf_gamma}
\gamma_m=\sigma\!\left(f_\gamma\!\left(
[v_m,s_m,v_m-s_m,v_m\odot s_m]+e_m^\gamma
\right)\right).
\end{equation}
The learned embedding $e_m^\gamma$ identifies the modality in the shared gate MLP. The adjusted summary and final prediction are:
\begin{equation}
\label{eq:msrgf_prediction}
\begin{aligned}
u_m&=v_m-\gamma_m\odot s_m,\\
\hat y&=f_{\mathrm{pred}}([c,u_T,u_A,u_V]).
\end{aligned}
\end{equation}
Thus, the MLP $f_{\mathrm{pred}}$ receives the joint context alongside the three adjusted modality summaries and predicts the sample-level sentiment $\hat y$. The global context provides a joint representation for sentiment prediction and a reference for adjusting individual modality summaries. The gated subtraction allows each modality summary to be modified according to the joint context, while the final predictor retains direct access to $c$.

\subsection{Learning Objectives}\label{sec:msrgf_losses}
The reconstruction and fusion stages are supervised at the representations they produce. In controlled missingness experiments, $X_m^\star$ is the complete training target and $\Omega_m=P_m\odot(1-M_m)$ identifies missing valid positions. Let $\ell(U,V;W)$ denote masked $L_1$ distance, averaged over selected positions and feature dimensions. The notation $\langle\cdot\rangle_{\Omega}$ averages over sample--modality pairs with at least one missing valid position, while $\langle\cdot\rangle_M$ averages over pairs with observed positions; within a pair, source errors are averaged over available modalities.

SWR receives direct supervision for every separate estimate, encouraging estimates from different available modalities to reconstruct the same complete features:
\begin{equation}
\label{eq:msrgf_view_loss}
\mathcal L_{\mathrm{view}}=
\left\langle\frac{1}{|\mathcal O|}\sum_{n\in\mathcal O}
\ell(D_m(R_m^n),X_m^\star;\Omega_m)\right\rangle_{\Omega}.
\end{equation}
RGF receives reconstruction supervision after the estimates have been fused:
\begin{equation}
\label{eq:msrgf_rec_loss}
\mathcal L_{\mathrm{rec}}=
\left\langle\ell(D_m(\widetilde H_m),X_m^\star;\Omega_m)\right\rangle_{\Omega}.
\end{equation}
Finally, to limit changes at observed positions, we encourage the enhanced representation to remain close to the projected observed features:
\begin{equation}
\label{eq:msrgf_pres_loss}
\mathcal L_{\mathrm{pres}}=
\left\langle\ell(\widetilde H_m,Z_m;M_m)\right\rangle_M.
\end{equation}
Here $Z_m$ is trainable, so this term encourages consistency without enforcing exact preservation. The sentiment objective $\mathcal L_{\mathrm{task}}$ is $L_1$ regression for CMU-MOSI and CMU-MOSEI, and the mean cross-entropy of four binary emotion classifiers for IEMOCAP. We optimize
\begin{equation}
\label{eq:msrgf_total}
\mathcal L=\mathcal L_{\mathrm{task}}
+\lambda_{rec}\mathcal L_{\mathrm{rec}}
+\lambda_{view}\mathcal L_{\mathrm{view}}
+\lambda_{pres}\mathcal L_{\mathrm{pres}}.
\end{equation}

%% file: 4_experiments.tex
\section{Experiments}
\label{sec:msrgf_experiments}

\subsection{Datasets and Evaluation Metrics}
We evaluate MSRGF on three benchmark datasets: CMU-MOSI \cite{zadeh2016mosi}, CMU-MOSEI \cite{zadeh2018multimodal}, and IEMOCAP \cite{busso2008iemocap}. CMU-MOSI and CMU-MOSEI contain 2,199 and 22,856 utterances, respectively, with sentiment scores ranging from $-3$ to $+3$. Following prior work \cite{zhuang2026hyper}, we convert sentiment predictions and labels into binary classes (positive and negative) and report accuracy (Acc) and F1-score, excluding neutral samples. IEMOCAP contains 4,453 samples across four binary emotion tasks (happy, sad, angry, and neutral), for which we report mean accuracy and F1-score.

\subsection{Implementation Details}

\textbf{Feature Extraction.}
We employ DeBERTa \cite{he2021debertav3,he2021deberta} for textual features on CMU-MOSI and CMU-MOSEI, and GloVe \cite{pennington2014glove} for IEMOCAP. For all three datasets, visual and acoustic features are extracted using OpenFace \cite{baltruvsaitis2016openface} and COVAREP \cite{degottex2014covarep}, respectively.

\textbf{Missingness Protocols.}
We consider three missingness settings. (1) \emph{Fixed whole-modality missingness} \cite{li2024corrkd,zhuang2025cmad} evaluates seven fixed-modality combinations: T, A, V, TA, TV, AV, and TAV, where TAV represents complete input. (2) \emph{Random whole-modality missingness} \cite{zhuang2026hyper,lian2023gcnet} randomly removes entire modalities at rates ranging from 10\% to 70\%, while ensuring that at least one modality remains available per sample. (3) \emph{Random within-modality missingness} \cite{zou2026dear,li2026semmsa,li2025tfmamba} randomly removes temporal features within each modality at rates ranging from 10\% to 90\%.

\textbf{Experimental Setup.}
All models are implemented in PyTorch \cite{paszke2017automatic} and trained on an NVIDIA RTX A6000 GPU. We set the hidden dimension to 128 and employ the AdamW optimizer \cite{loshchilov2018decoupled}. Additional implementation details are provided in the Appendix.

\begin{table*}[t]
\centering
\caption{Comparison under fixed whole-modality missingness (Acc/F1, \%). Bold and underline indicate the best and second-best results, with ties sharing rank. $^\ast$ marks results reported by HME~\cite{zhuang2026hyper}; $^\dagger$ denotes averages over six available combinations.}
\label{tab:msrgf_fixed_all}
\renewcommand{\arraystretch}{1.10}
\small
\setlength{\tabcolsep}{4pt}
\begin{tabular*}{\textwidth}{@{\extracolsep{\fill}}lcccccccc@{}}
\toprule
Method & T & A & V & TA & TV & AV & TAV & Avg \\
\midrule
\multicolumn{9}{@{}l}{\textbf{CMU-MOSI}} \\
DiCMoR$^\ast$~\cite{wang2023dicmor} & 84.5/84.4 & 60.5/60.8 & 62.2/60.2 & 85.5/85.5 & 85.5/85.4 & 64.0/63.5 & 85.7/85.6 & 75.4/75.1 \\
IMDer$^\ast$~\cite{wang2023imder} & 84.8/84.7 & 62.0/62.2 & 61.3/60.8 & 85.4/85.3 & 85.5/85.4 & 63.6/63.4 & 85.7/85.6 & 75.5/75.3 \\
LNLN$^\ast$~\cite{zhang2024lnln} & 84.9/85.1 & 52.2/58.9 & 52.2/58.9 & 84.9/85.2 & 84.9/85.1 & 52.2/58.9 & 84.3/84.6 & 70.8/73.8 \\
MPLMM$^\ast$~\cite{guo2024mplmm} & 83.8/83.8 & 54.6/42.0 & 54.7/42.1 & 83.7/83.6 & 83.8/83.7 & 57.3/42.6 & 83.8/83.8 & 71.7/65.9 \\
HME$^\ast$~\cite{zhuang2026hyper} & 85.7/85.6 & 62.7/62.2 & \underline{63.6}/\underline{63.4} & 85.8/85.7 & 85.8/85.8 & 64.6/64.5 & \underline{86.4}/\underline{86.3} & \underline{76.4}/\underline{76.2} \\
API~\cite{wang2026api} & \underline{85.8}/\underline{85.8} & \underline{63.3}/\textbf{63.2} & \textbf{64.5}/\textbf{64.4} & \underline{86.3}/\underline{86.1} & \underline{86.1}/\underline{86.0} & \underline{65.2}/\textbf{65.2} & -- & 75.2/75.1$^\dagger$ \\
TLRA~\cite{chen2026tlra} & 85.2/85.2 & 62.5/62.0 & 62.7/62.2 & 85.8/86.0 & 85.5/85.5 & 61.3/61.3 & 86.3/86.2 & 75.6/75.5 \\
\midrule
\textbf{MSRGF} & \textbf{86.9}/\textbf{86.8} & \textbf{63.6}/\underline{62.9} & 63.0/61.9 & \textbf{86.4}/\textbf{86.5} & \textbf{86.6}/\textbf{86.6} & \textbf{65.7}/\underline{64.9} & \textbf{89.8}/\textbf{89.8} & \textbf{77.4}/\textbf{77.1} \\
\midrule
\multicolumn{9}{@{}l}{\textbf{CMU-MOSEI}} \\
DiCMoR$^\ast$~\cite{wang2023dicmor} & 84.2/84.3 & 62.9/60.4 & 63.6/63.6 & 85.0/84.9 & 84.9/84.9 & 65.2/64.4 & 85.1/85.1 & 75.8/75.4 \\
IMDer$^\ast$~\cite{wang2023imder} & 84.5/84.5 & 63.8/60.6 & 63.9/63.6 & 85.1/85.1 & 85.0/85.0 & 64.9/63.5 & 85.1/85.1 & 76.0/75.3 \\
MPLMM$^\ast$~\cite{guo2024mplmm} & 84.6/84.6 & 61.8/49.4 & 62.2/53.2 & 85.1/85.1 & 85.2/85.1 & 63.2/51.0 & 85.3/85.3 & 75.3/70.5 \\
HME$^\ast$~\cite{zhuang2026hyper} & 85.5/85.4 & 64.1/59.7 & 64.4/62.9 & 85.6/85.4 & 85.7/85.7 & 65.4/63.7 & \underline{86.2}/\underline{86.2} & 76.7/75.6 \\
CMAD~\cite{zhuang2025cmad} & \underline{86.1}/\underline{86.0} & 63.0/60.8 & 65.7/64.4 & \underline{86.3}/\underline{86.2} & \underline{86.4}/\underline{86.4} & 65.6/64.8 & 86.1/\underline{86.2} & 77.0/76.4 \\
API~\cite{wang2026api} & 85.5/85.4 & 65.1/\textbf{65.5} & \textbf{66.7}/\underline{65.4} & 85.7/85.5 & 85.6/85.6 & 66.1/\underline{66.0} & -- & 75.8/75.6$^\dagger$ \\
TLRA~\cite{chen2026tlra} & 85.1/85.1 & \textbf{65.5}/63.7 & 66.0/\textbf{65.5} & 85.9/85.8 & 84.8/84.8 & \underline{66.6}/65.6 & 85.9/85.9 & \underline{77.1}/\underline{76.6} \\
\midrule
\textbf{MSRGF} & \textbf{87.5}/\textbf{87.5} & \underline{65.2}/\underline{64.2} & \underline{66.4}/\textbf{65.5} & \textbf{87.1}/\textbf{87.1} & \textbf{87.2}/\textbf{87.3} & \textbf{67.3}/\textbf{66.1} & \textbf{88.2}/\textbf{88.1} & \textbf{78.4}/\textbf{78.0} \\
\midrule
\multicolumn{9}{@{}l}{\textbf{IEMOCAP}} \\
DiCMoR$^\ast$~\cite{wang2023dicmor} & 75.2/68.2 & 76.0/68.7 & 73.6/68.8 & 78.7/76.5 & 74.4/72.1 & 76.3/75.2 & 78.7/76.5 & 76.1/72.3 \\
IMDer$^\ast$~\cite{wang2023imder} & 75.1/67.2 & 75.5/70.2 & 75.1/70.1 & 76.6/71.7 & 75.2/72.6 & 76.3/72.3 & 77.5/75.4 & 75.9/71.4 \\
LNLN$^\ast$~\cite{zhang2024lnln} & 76.8/74.4 & 77.1/74.6 & 74.5/69.9 & 77.9/75.4 & 77.0/75.5 & 77.3/74.9 & 78.4/77.4 & 77.0/74.6 \\
MPLMM$^\ast$~\cite{guo2024mplmm} & 69.2/69.3 & 59.8/59.7 & 57.6/57.0 & 76.0/75.4 & 74.7/74.5 & 67.3/67.4 & -- & 67.4/67.2$^\dagger$ \\
HME$^\ast$~\cite{zhuang2026hyper} & 79.1/78.1 & 78.9/\underline{76.8} & \underline{75.5}/\underline{70.8} & 80.7/80.0 & \underline{79.7}/78.9 & 78.1/76.3 & \underline{81.1}/80.6 & 79.0/77.4 \\
CMAD~\cite{zhuang2025cmad} & \textbf{79.5}/\textbf{78.7} & \underline{79.2}/76.7 & 73.6/70.6 & \textbf{81.9}/\textbf{81.3} & \underline{79.7}/\underline{79.1} & \underline{79.0}/\underline{77.3} & \textbf{81.7}/\textbf{81.2} & \underline{79.2}/\underline{77.8} \\
\midrule
\textbf{MSRGF} & \underline{79.3}/\underline{78.6} & \textbf{79.7}/\textbf{77.7} & \textbf{75.6}/\textbf{71.4} & \underline{81.2}/\underline{80.6} & \textbf{80.9}/\textbf{80.6} & \textbf{79.7}/\textbf{77.6} & \textbf{81.7}/\underline{80.8} & \textbf{79.7}/\textbf{78.2} \\
\bottomrule
\end{tabular*}
\end{table*}

\begin{figure*}[t]
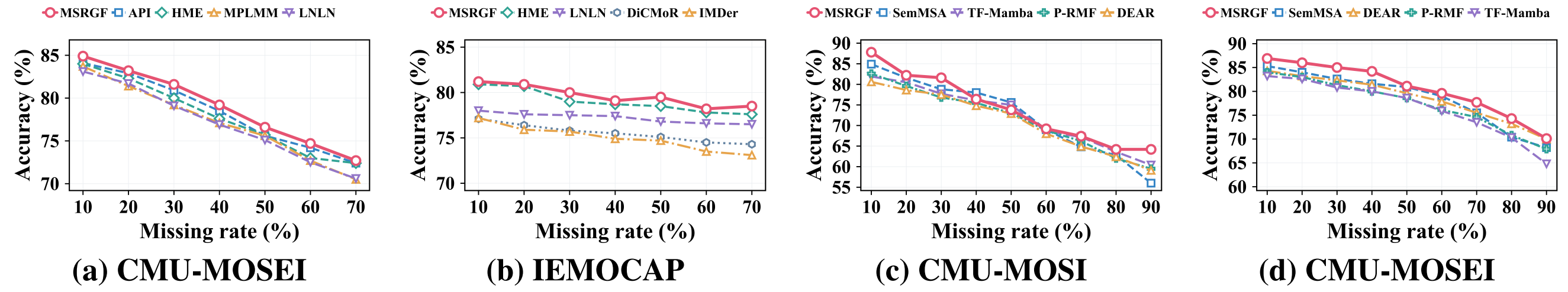

    \centering
    \begin{minipage}[t]{0.24\textwidth}
        \centering
        \includegraphics[width=\linewidth]{fig_plot_1.png}\\[-1mm]
        \textbf{(a) CMU-MOSEI}
    \end{minipage}\hfill
    \begin{minipage}[t]{0.24\textwidth}
        \centering
        \includegraphics[width=\linewidth]{fig_plot_2.png}\\[-1mm]
        \textbf{(b) IEMOCAP}
    \end{minipage}\hfill
    \begin{minipage}[t]{0.24\textwidth}
        \centering
        \includegraphics[width=\linewidth]{fig_plot_3.png}\\[-1mm]
        \textbf{(c) CMU-MOSI}
    \end{minipage}\hfill
    \begin{minipage}[t]{0.24\textwidth}
        \centering
        \includegraphics[width=\linewidth]{fig_plot_4.png}\\[-1mm]
        \textbf{(d) CMU-MOSEI}
    \end{minipage}
    \caption{Accuracy performance under random missingness. Figures (a,b) show random whole-modality missingness with missing rates from 10\% to 70\%; Figures (c,d) show random within-modality missingness with missing rates from 10\% to 90\%. }
    \label{fig:msrgf_acc_overview}
\end{figure*}

\begin{figure}[t]
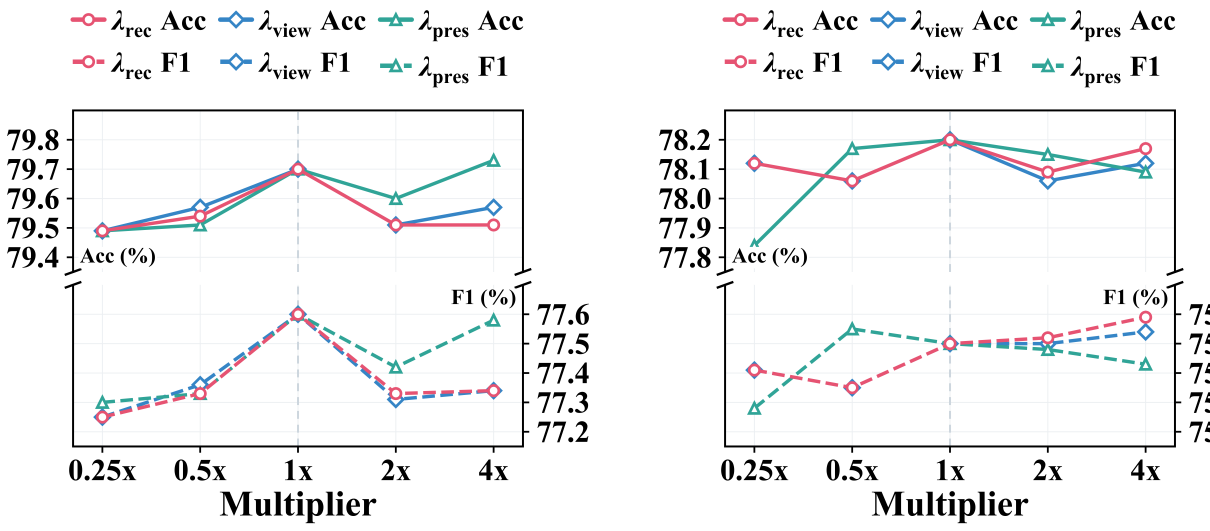

    \centering
    \begin{minipage}[t]{0.23\textwidth}
        \centering
        \includegraphics[width=\linewidth]{mdrgf_loss_hy_av.png}\\[-1mm]
        \textbf{(a) AV}
    \end{minipage}\hfill
    \begin{minipage}[t]{0.23\textwidth}
        \centering
        \includegraphics[width=\linewidth]{mdrgf_loss_hy_random60.png}\\[-1mm]
        \textbf{(b) MR=60\%}
    \end{minipage}\hfill
    \caption{Loss weight sensitivity on IEMOCAP under (a) AV and (b) random whole-modality missingness (MR=60\%). Multipliers scale each default weight $(0.1,0.1,0.05)$ independently.}
    \label{fig:sss}
\end{figure}

\subsection{Comparison with State-of-the-Art Methods}
We compare MSRGF with representative incomplete MSA methods, including DiCMoR \cite{wang2023dicmor}, IMDer \cite{wang2023imder}, LNLN \cite{zhang2024lnln}, MPLMM \cite{guo2024mplmm}, HME \cite{zhuang2026hyper}, CMAD \cite{zhuang2025cmad}, API \cite{wang2026api}, and TLRA \cite{chen2026tlra}. For random within-modality missingness, we additionally include ALMT \cite{zhang2023almt}, P-RMF \cite{zhu2025prmf}, DEAR \cite{zou2026dear}, SemMSA \cite{li2026semmsa}, and TF-Mamba \cite{li2025tfmamba}. Detailed results are provided in the Appendix.

\textbf{Fixed Whole-Modality Missingness.} As shown in Table~\ref{tab:msrgf_fixed_all}, MSRGF achieves the best average Acc/F1 across all three datasets, reaching 77.4/77.1 on MOSI, 78.4/78.0 on MOSEI, and 79.7/78.2 on IEMOCAP. Compared with the strongest baselines, these results yield improvements of 1.0/0.9, 1.3/1.4, and 0.5/0.4 percentage points, respectively, demonstrating its effectiveness across different available-modality combinations.

\textbf{Random Missingness.} Figure~\ref{fig:msrgf_acc_overview}(a,b) presents results under random whole-modality missingness (10\%--70\%), where MSRGF achieves average Acc/F1 scores of 79.0/78.3 on MOSEI and 79.6/78.2 on IEMOCAP, outperforming the strongest baselines by 0.6/0.5 and 0.6/0.4 points, respectively. Under random within-modality missingness, Figure~\ref{fig:msrgf_acc_overview}(c,d) shows that MSRGF achieves 74.1/74.1 on MOSI and 80.5/80.2 on MOSEI, improving over the strongest baselines by 1.7/1.6 and 1.9/1.0 points, respectively. These results demonstrate the robustness of MSRGF across different missingness types and rates.

\subsection{Ablation Studies}\label{ablas_in_main}
We evaluate the contributions of SWR, RGF, and GCM by averaging results over nine random within-modality missing rates on MOSI and MOSEI, and seven random whole-modality missing rates on IEMOCAP. As shown in Table~\ref{tab:msrgf_ablation}, removing SWR causes F1 reductions of 2.9, 2.2, and 3.0 percentage points, respectively, highlighting the importance of source-wise reconstruction. Replacing RGF with uniform completion reduces F1 by 1.0--2.9 points, supporting the benefit of reference-guided integration over direct averaging of source estimates. Removing GCM also decreases F1 by 1.1--2.6 points, demonstrating the contribution of global context modulation. Additional ablations on IEMOCAP further examine the internal mechanisms and auxiliary objectives under fixed and random whole-modality missingness, as detailed in Appendix~\ref{sec:msrgf_supp_mech} (Table~\ref{tab:msrgf_ablation_add}).

\subsection{Hyperparameter Analysis}

We also examine the sensitivity of three loss coefficients, $\lambda_{\mathrm{rec}}$, $\lambda_{\mathrm{view}}$, and $\lambda_{\mathrm{pres}}$, on IEMOCAP under the AV combination and random whole-modality missingness at MR=60\%. As shown in Figure~\ref{fig:sss}, each coefficient is independently scaled from 0.25 to 4 times its default value, with the others fixed. Overall, performance exhibits relatively small fluctuations across different multipliers, with the default configuration providing consistently competitive results on both settings. Although some alternative weights yield marginal improvements, no consistent benefit is observed from increasing or decreasing individual coefficients, suggesting that MSRGF is relatively insensitive to these loss weights within the evaluated range.

\begin{table}[t]
\centering
\caption{Component ablations, averaged over nine within-modality missing rates on MOSI/MOSEI and seven random whole-modality missing rates on IEMOCAP.}
\label{tab:msrgf_ablation}
\setlength{\tabcolsep}{2.5pt}
\renewcommand{\arraystretch}{1.17}
\begin{tabular*}{\columnwidth}{
@{\extracolsep{\fill}}lccc@{}}
\toprule
Method & MOSI & MOSEI & IEMOCAP\\
\midrule
w/o SWR & 71.2/71.2 & 77.9/78.0 & 77.6/75.2 \\
w/o RGF & 73.1/73.1 & 78.9/78.1 & 77.7/75.3 \\
w/o GCM & 72.8/73.0 & 79.4/79.0 & 77.9/75.6 \\
\midrule
\textbf{MSRGF}
& \textbf{74.1/74.1} & \textbf{80.5/80.2} & \textbf{79.6/78.2} \\
\bottomrule
\end{tabular*}
\end{table}

\subsection{Mechanism Analysis}\label{sec:msrgf_mechanism_analysis}
We investigate the roles of source-wise estimates and RGF through inference interventions and retrained design alternatives. 

\begin{table}[t]
\centering
\caption{Inference interventions on IEMOCAP (F1, \%).}
\label{tab:msrgf_mechanism_main}
\setlength{\tabcolsep}{2pt}
\renewcommand{\arraystretch}{1.10}
\begin{tabular*}{\columnwidth}{@{\extracolsep{\fill}}lccccc@{}}
\toprule
 & & SWR & \multicolumn{3}{c}{RGF update} \\
\cmidrule(lr){3-3}\cmidrule(l){4-6}
Setting & Full & Single & Zero & Random & Perm. \\
\midrule
TV$\rightarrow$A & \textbf{80.61} & 79.20 & 80.00 & \underline{80.03} & 79.60 \\
AV$\rightarrow$T & \textbf{77.58} & 77.03 & \underline{77.33} & 77.28 & 76.88 \\
\bottomrule
\end{tabular*}
\end{table}

\textbf{Source-wise estimates.} We first examine the contribution of multiple reconstruction sources through inference interventions and retrained design alternatives. As shown in Table~\ref{tab:msrgf_mechanism_main}, retaining both source estimates improves F1 by 1.41 and 0.55 points under TV and AV, respectively, compared with the stronger tested single source. We further compare MSRGF with retrained T-, A-, and V-priority single-source variants and a Fused-source Generator. All alternatives yield lower average Acc/F1 under fixed and random whole-modality missingness (Appendix\ref{sec:msrgf_supp_mech}, Table~\ref{tab:msrgf_ablation_add}). These results demonstrate the benefit of retaining separate source-conditioned estimates for subsequent reference-guided fusion.

\textbf{Effective difference update.} We then assess the learned update $U_m=g_m^{\mathrm{enh}}\odot N_m$ by replacing it with zero, an equal-norm random direction, or an update permuted across samples. Only the missing target is modified, while other enhanced representations remain unchanged. The original update outperforms all alternatives under TV and AV (Table~\ref{tab:msrgf_mechanism_main}), indicating that the learned refinement contributes to prediction.

\begin{table}[t]
\centering
\caption{Retrained RGF ablations on IEMOCAP (Acc/F1, \%).}
\label{tab:msrgf_rgf_alternatives}
\setlength{\tabcolsep}{3pt}
\renewcommand{\arraystretch}{1.10}
\begin{tabular*}{\columnwidth}{@{\extracolsep{\fill}}lcc@{}}
\toprule
Variant & Fixed & Random \\
\midrule
Uniform $\alpha$ & 78.38/76.12 & 78.05/75.74 \\
Early Aggregation & \underline{78.43}/76.16 & 78.05/75.74 \\
Ordinary Residual & 78.30/76.02 & 78.03/75.75 \\
w/o Difference Refinement & 78.34/76.06 & 78.05/75.72 \\
\addlinespace[2pt]
Hard Observation Copy & 77.96/75.76 & 77.83/75.38 \\
w/o Observed Fusion & 78.40/\underline{76.79} & \underline{78.13}/\underline{76.35} \\
\midrule
\textbf{MSRGF} & \textbf{79.73}/\textbf{78.17} & \textbf{79.63}/\textbf{78.25} \\
\bottomrule
\end{tabular*}
\end{table}

\textbf{Reference construction and difference refinement.} We next evaluate RGF against simpler fusion strategies. Uniform $\alpha$ averages source estimates with equal weights, while Early Aggregation merges them before RGF. Ordinary Residual replaces source-specific differences with $C_m-B_m$, and w/o Difference Refinement removes $N_m$. All variants yield lower Acc/F1 than MSRGF under both protocols (Table~\ref{tab:msrgf_rgf_alternatives}), demonstrating the advantage of learned reference construction and source-specific refinement.

\textbf{Observed-feature integration.} Finally, we examine how RGF incorporates retained observations. Hard Observation Copy replaces enhanced features at observed positions with $Z_m$, whereas w/o Observed Fusion removes the direct contribution of $B_m$ and its interpolation toward $C_m$. Both variants degrade performance, reducing F1 by 2.41/2.87 and 1.38/1.90 points under Fixed/Random, respectively. This highlights the importance of integrating observed and reconstructed information rather than relying on strict copying or removing the observation fusion pathway.

Additional ablations and mechanism analyses, including source-wise estimate contributions, reconstruction diagnostics, relation and refinement interventions, observed-feature preservation, and global context modulation, are provided in Appendix~\ref{sec:msrgf_supp_mech}.

%% file: 5_conclusion.tex
\section{Conclusion}

In this paper, we proposed MSRGF for incomplete multimodal sentiment analysis, addressing the insufficient exploitation of complementary information among source-wise estimates. Specifically, Source-Wise Reconstruction generates separate target estimates from available modalities through a shared reconstruction generator. Reference-Guided Fusion integrates these estimates by constructing a weighted reference and refining source-specific differences, while incorporating retained observations to enhance target representations. Global Context Modulation further adjusts the enhanced representations using global context for sentiment prediction. Extensive experiments on CMU-MOSI, CMU-MOSEI, and IEMOCAP demonstrate the effectiveness of MSRGF across diverse missingness.

\section*{LLM Usage Statement}
During the preparation of this work, the authors utilized OpenAI's ChatGPT for language refinement. AI tools were additionally employed in generating and improving implementation code, as well as facilitating discussions on research directions and experimental procedures. The authors reviewed and revised all AI-assisted materials and take full responsibility for the content, accuracy, and integrity of this work.

%% file: X_suppl.tex
\clearpage
\setcounter{page}{1}
\maketitlesupplementary

\section{Additional Method Details}\label{sec:msrgf_supp_method}
This section provides the implementation details of MSRGF, following the computation from masked multimodal inputs to the final prediction. The missingness protocol determines the observed features available to Source-Wise Reconstruction (SWR), whose target estimates are subsequently combined by Reference-Guided Fusion (RGF). Global Context Modulation (GCM) then operates on the enhanced modality representations, while the network configurations at the end of the section specify the corresponding dimensions and parameter-sharing rules.

\subsection{Input Features and Missingness}\label{sec:msrgf_supp_inputs}
MSRGF operates on aligned textual, acoustic, and visual features, with observation masks specifying the inputs used by SWR and the positions retained during RGF integration. The input dimensions $(d_T,d_A,d_V)$ are $(1024,5,20)$ for CMU-MOSI, $(1024,74,35)$ for CMU-MOSEI, and $(300,74,35)$ for IEMOCAP. The common hidden dimension is $d=128$. CMU-MOSI and CMU-MOSEI use text features; words selected for masking are replaced by \texttt{[UNK]} before contextual encoding, and their word-level features are excluded from the observed input. IEMOCAP uses pre-extracted features for all three modalities. Complete features $X_m^\star$ provide reconstruction supervision during training, while the prediction network receives $X_m^{\mathrm{obs}}$, $M_m$, and $P_m$.

To distinguish artificial masking from positions without valid supervision, let $A_m$ indicate positions with an available reconstruction target before masking. Together with the observation mask $M_m$ and validity mask $P_m$ defined in the main paper, it satisfies $M_m\leq A_m\leq P_m$. Reconstruction supervision is therefore applied at:
\begin{equation}
\label{eq:msrgf_supp_positions}
\Omega_m=A_m\odot(1-M_m).
\end{equation}

For the controlled missingness protocols, $A_m=P_m$ and hence $\Omega_m=P_m\odot(1-M_m)$. Even when a modality is entirely absent, its aligned sequence length and validity mask are preserved, allowing SWR to construct target estimates at the same temporal positions as in the complete sequence.

The fixed-modality setting retains each nonempty subset of $\{T,A,V\}$. For random whole-modality missingness, a modality is removed in full, with at least one modality retained per sample. We evaluate missing rates from $10\%$ to $70\%$ in steps of $10\%$; the maximum attainable rate with three modalities is $2/3$, which is used for the $70\%$ condition. For random within-modality missingness, temporal positions are randomly masked at rates from $10\%$ to $90\%$ over the valid sequence.

\subsection{Source-Wise Reconstruction}\label{sec:msrgf_supp_swr}
Once the available modalities have been determined by the masks, SWR constructs a separate estimate $R_m^n$ of target modality $m$ from each observed source $n\in\mathcal O$. Each path combines a target-specific query with the encoded observations of its source, allowing a shared generator to handle different source--target pairs. Specifically, the observed features are first mapped to $Z_n$ through a modality-specific linear projection followed by LayerNorm (together denoted by $\phi_n$), then encoded as $H_n$ by a one-layer temporal Transformer. The Transformer uses four attention heads, a $128\!\to\!256\!\to\!128$ feed-forward network with GELU, and dropout $0.1$. Its attention reads the positions marked by $M_n$; an entirely missing modality yields $H_n=\mathbf0$.

To generate an estimate at each target position, SWR forms $q_m^0(t)$ from the positional, modality, and observation-state embeddings defined in the main paper. A target-specific prompt bank $U_m\in\mathbb R^{4\times128}$ further adjusts this query through attention over four learned vectors:
\begin{equation}
\label{eq:msrgf_supp_prompt}
\begin{aligned}
p_m(t)&=\operatorname{softmax}\!\left(\frac{q_m^0(t)U_m^\top}{\sqrt d}\right)U_m,\\
Q_m(t)&=P_m(t)f_Q\big(q_m^0(t)+p_m(t)\big).
\end{aligned}
\end{equation}
Here $f_Q$ is the shared $128\!\to\!128\!\to\!128$ GELU MLP that maps the prompt-adjusted query to $Q_m$, which is then reused in the SWR generator and the contextual base construction of RGF. The learned positional embedding table covers 512 positions, whereas the modality and observation-state embeddings contain three and two entries, respectively.

For each $n\in\mathcal O$, the shared generator reads $H_n$ using $Q_m$. It contains two consecutive cross-attention blocks. For query features $q$, encoded features $h$, and observation mask $M$, each block computes:
\begin{equation}
\label{eq:msrgf_supp_cross_attn}
\begin{aligned}
J&=\operatorname{MHA}\big(\operatorname{LN}_q(q),
\operatorname{LN}_h(h),\operatorname{LN}_h(h);M\big),\\
x&=q+\operatorname{Dropout}(J),\\
q'&=x+\operatorname{Dropout}\big(\operatorname{FFN}(\operatorname{LN}(x))\big).
\end{aligned}
\end{equation}

Each cross-attention block uses separate normalization layers for queries and observed features, four attention heads, and a width-256 GELU feed-forward network. Although the two blocks have independent parameters, both are reused across source--target pairs. If a cross-attention call receives no observed keys, its attention update is zero while the residual and feed-forward operations remain defined; source paths without observed positions are excluded from the set of reconstructed estimates.

The generator output is mapped by a modality-specific two-layer GELU MLP $f_m^{\mathrm{tar}}$ with widths $128\!\to\!128\!\to\!128$. The generator is shared across modality pairs, and the same output head is used for all available modalities estimating a fixed $m$. Modality embeddings are added to the generator's query and observed-feature inputs. The resulting estimates $R_m^n\in\mathbb R^{L\times128}$ share the target's temporal positions. For each modality, a separate linear decoder $D_m:\mathbb R^{128}\to\mathbb R^{d_m}$ (with bias) is shared by its source-wise estimates and final enhanced representation. The same decoder is used for reconstruction supervision and the observed-feature comparison in RGF, including at inference when this comparison is defined.

\subsection{Reference-Guided Fusion}\label{sec:msrgf_supp_rgf}
The separate estimates produced by SWR provide multiple candidate representations for a target modality, but their contributions must be determined in relation to the observed inputs and to one another. RGF first constructs an observation-conditioned base $B_m$ and relation descriptors $\xi_m^n$, which guide the learned reference $C_m$. It then transforms the deviations of individual estimates from this reference into $N_m$ before integrating both terms into $\widetilde H_m$. The same observation mask controls how this integration behaves at observed and missing positions.

\textbf{Base representation.} RGF begins with the base representation $B_m$, which preserves a direct contribution from $Z_m$ wherever target observations exist. A cross-attention block, parameterized separately from the SWR generator, reads $H_m$ using the target query $Q_m$; its output passes through the shared $128\!\to\!128\!\to\!128$ GELU MLP $f_{\mathrm{anc}}$ to obtain $a_m$. With $\eta_a=0.1$, the base is defined by:
\begin{equation}
\label{eq:msrgf_supp_base_cases}
\begin{gathered}
B_m(t)=\\[-2pt]
\begin{cases}
Z_m(t)+\eta_a a_m(t),&M_m(t)=1,\\
a_m(t),&M_m(t)=0,\ P_m(t)=1,\\
\mathbf0,&P_m(t)=0.
\end{cases}
\end{gathered}
\end{equation}
This definition gives observed positions a direct path from $Z_m$ while providing a contextual base at missing positions. When the entire target modality is absent, $a_m$ is obtained through the query residual and feed-forward operations of the base block. Because $B_m$ also enters the reference scorer, refinement scorer, and integration gates, the reconstruction paths are evaluated against a common target-conditioned representation.

\textbf{Relations among estimates.} Using $B_m$ alone does not distinguish estimates produced from different sources. RGF therefore constructs $\xi_m^n$ from three complementary comparisons: the relation of $R_m^n$ to its source $n$, its variation from the other estimates of target $m$, and its agreement with the retained target observations where available. Three indicators specify which comparisons are valid:
\begin{equation}
\label{eq:msrgf_supp_rel_validity}
\begin{aligned}
v_{m,n}^{\mathrm{tar}} &= \mathbb I[\|M_m\|_1>0,\ n\ne m],\\
v_n^{\mathrm{src}} &= \mathbb I[n\in\mathcal O],\\
v_n^{\mathrm{div}} &= \mathbb I[|\mathcal O|>1].
\end{aligned}
\end{equation}
Here $v_{m,n}^{\mathrm{tar}}$ enables agreement with retained target observations, $v_n^{\mathrm{src}}$ identifies available source slots, and $v_n^{\mathrm{div}}$ permits comparison across distinct source estimates. As descriptors are computed only for $n\in\mathcal O$, the source indicator equals one on active paths but is kept for the fixed-slot implementation. The source--estimate interaction starts from masked pooled representations:
\begin{equation}
\label{eq:msrgf_supp_masked_pool}
\bar R_m^n=\operatorname{Pool}(R_m^n,P_m),\quad
\bar H_n=\operatorname{Pool}(H_n,M_n),
\end{equation}
where:
\begin{equation}
\operatorname{Pool}(X,W)=\sum_t W(t)X(t)/\max(\sum_t W(t),1).
\end{equation}
These source and target summaries are then projected to 64 dimensions and combined with learned modality embeddings:
\begin{equation}
\label{eq:msrgf_supp_rel_proj}
\begin{aligned}
a_m^n&=\psi_m(\bar R_m^n)+e_m^{\mathrm{sem}},\\
b_n&=\psi_n(\bar H_n)+e_n^{\mathrm{sem}}.
\end{aligned}
\end{equation}

A shared MLP processes their values, element-wise product, and absolute difference to obtain:
\begin{equation}
\label{eq:msrgf_supp_rel_source}
\mathbf c_{m,n}^{\mathrm{src}}=f_{\mathrm{src}}\big(
[a_m^n,b_n,a_m^n\odot b_n,|a_m^n-b_n|]\big).
\end{equation}
The network $f_{\mathrm{src}}$ has widths $256\!\to\!64\!\to\!16$, producing a 16-dimensional interaction feature.

The second comparison measures the difference between $R_m^n$ and the other estimates of modality $m$. The corresponding score is:
\begin{equation}
\label{eq:msrgf_supp_rel_div}
\begin{aligned}
c_{m,n}^{\mathrm{div}}={}&v_n^{\mathrm{div}}\operatorname{sg}\Bigg[
\frac{1}{\max(|\mathcal O|-1,1)}\\
&\cdot\sum_{j\in\mathcal O\setminus\{n\}}
\big(1-\cos(\bar R_m^n,\bar R_m^j)\big)\Bigg].
\end{aligned}
\end{equation}
where $\cos$ is cosine similarity, $\operatorname{sg}$ stops gradients through the score, and the availability indicator is defined in Equation~\eqref{eq:msrgf_supp_rel_validity}. This comparison uses the estimates obtained from different available modalities for the same $m$.

When modality $m$ contains observed positions and $n\ne m$, the third comparison measures agreement between the decoded estimate and the observed features:
\begin{equation}
\label{eq:msrgf_supp_rel_target}
\begin{aligned}
c_{m,n}^{\mathrm{tar}}={}&v_{m,n}^{\mathrm{tar}}\operatorname{sg}\Bigg[\\
&\frac{\sum_t M_m(t)\cos\big(D_m(R_m^n(t)),X_m^{\mathrm{obs}}(t)\big)}
{\max(\sum_t M_m(t),1)}\Bigg].
\end{aligned}
\end{equation}

The availability condition is given in Equation~\eqref{eq:msrgf_supp_rel_validity}; the cosine score is computed without gradients. Observation coverage further describes the amount of information available in each modality:
\begin{equation}
\label{eq:msrgf_supp_coverage}
\rho_m=\frac{\|M_m\|_1}{\max(\|P_m\|_1,1)}.
\end{equation}
Combining these measurements and their validity indicators gives:
\begin{equation}
\label{eq:msrgf_supp_rel_vector}
\begin{aligned}
\xi_m^n=\big[&c_{m,n}^{\mathrm{tar}},\mathbf c_{m,n}^{\mathrm{src}},
c_{m,n}^{\mathrm{div}},\rho_m,\rho_n,\\
&v_{m,n}^{\mathrm{tar}},v_n^{\mathrm{src}},v_n^{\mathrm{div}}\big]
\in\mathbb R^{23},
\end{aligned}
\end{equation}

The resulting 23-dimensional $\xi_m^n$ combines the 16-dimensional source--estimate interaction with the two comparison scores, observation coverage, and validity indicators. It is computed once per available source and shared across the target positions, so the subsequent position-wise weighting networks can condition on both local features and source-level relations.

\textbf{Reference and refinement weights.} To turn the source-level comparisons into a position-wise reference, the reference scorer takes $R_m^n(t)$ together with $B_m(t)$ and $\xi_m^n$. After the reference is assembled, the feature-wise refinement gate additionally receives $C_m(t)$, allowing it to condition the source-specific differences on their shared reference. The corresponding inputs are:
\begin{equation}
\label{eq:msrgf_supp_score_inputs}
\begin{aligned}
I_m^n(t)&=[R_m^n(t),B_m(t),\xi_m^n]\in\mathbb R^{279},\\
J_m^n(t)&=[C_m(t),B_m(t),R_m^n(t),\xi_m^n]
\in\mathbb R^{407}.
\end{aligned}
\end{equation}
The reference weights $\alpha_m^n$ are computed from $I_m^n$ by a shared $279\!\to\!128\!\to\!1$ GELU MLP and normalized over available modalities. Their weighted sum of $R_m^n$ yields $C_m$ in Equation~\eqref{eq:msrgf_reference}. The refinement scorer uses a separate MLP of the same width to obtain $\beta_m^n$ from $I_m^n$, while a separate shared $407\!\to\!128\!\to\!128$ GELU MLP produces the feature-wise gates $G_m^n$ from $J_m^n$ with sigmoid activation. The scorer therefore determines how much each available modality contributes, and the gate controls the corresponding feature dimensions.

To retain modality identity while sharing the scoring networks across paths, the inputs include learned embeddings $e_m^\alpha\in\mathbb R^{279}$, $e_{m,n}^\beta\in\mathbb R^{279}$, and $e_{m,n}^G\in\mathbb R^{407}$, with indices corresponding to the target or source--target pair.

Once $C_m$ has been formed, each estimate is compared with the reference through $\Delta_m^n=R_m^n-C_m$. A shared bias-free $128\!\to\!256\!\to\!128$ GELU MLP $F_{\mathrm{diff}}$ transforms each difference before the source weights $\beta_m^n$ and feature gates $G_m^n$ are applied. Its two linear layers are bias-free, giving $F_{\mathrm{diff}}(\mathbf0)=\mathbf0$. The weighted transformed differences are summed to obtain $N_m$ in Equation~\eqref{eq:msrgf_difference}. $C_m$ defines the common reference, while $N_m$ combines the source-specific difference updates used in final integration.

\textbf{Final integration.} The reference and refinement are combined with the base representation using the observation mask. RGF first summarizes the relations as $\bar\xi_m(t)=\sum_{n\in\mathcal O}\alpha_m^n(t)\xi_m^n$, so the information used to construct $C_m$ also enters the integration gates. A shared $408\!\to\!128\!\to\!256$ GELU MLP followed by sigmoid receives $[B_m,C_m,N_m,\bar\xi_m,M_m]$ and produces two 128-dimensional gates $r_m^{\mathrm{rec}}$ and $r_m^{\mathrm{enh}}$. Their contributions are adjusted by $M_m$ according to Equation~\eqref{eq:msrgf_gates}, with $\eta_g=0.1$. This gives separate control over the reference update and the difference-based refinement at each position.

Substituting the observation-dependent gates into the RGF integration rule gives the following expression at an observed position:
\begin{equation}
\label{eq:msrgf_supp_final_observed}
\begin{aligned}
\widetilde H_m(t)={}&Z_m(t)+\eta_a a_m(t)\\
&+\eta_g r_m^{\mathrm{rec}}(t)\odot(C_m(t)-B_m(t))\\
&+\eta_g r_m^{\mathrm{enh}}(t)\odot N_m(t).
\end{aligned}
\end{equation}
At a valid missing position, the reference update has unit weight, reducing the expression to:
\begin{equation}
\label{eq:msrgf_supp_final_missing}
\widetilde H_m(t)=C_m(t)+r_m^{\mathrm{enh}}(t)\odot N_m(t).
\end{equation}
After applying $P_m$, padding positions remain zero. The two expressions expose the distinction central to RGF: observed features are retained through $B_m$ with controlled updates, whereas missing positions are represented by the learned reference and the transformed differences. Both paths use the same source relations and weighting networks.

\textbf{Available modalities.} The set $\mathcal O$ determines the estimates used in both the reference and refinement stages. When only one modality $n$ is available, its normalized reference weight is one, giving $C_m=R_m^n$ and $N_m=\mathbf0$. With two or three available modalities, each estimate contributes to $C_m$ through its learned weight, and the differences $R_m^n-C_m$ form the inputs to $N_m$. For instance, when text and audio are observed and vision is entirely missing, $R_V^T$ and $R_V^A$ are combined into $C_V$; their transformed differences then form $N_V$, and the two representations produce $\widetilde H_V$ through the missing-position gates. The same computations apply to partially observed modalities, with $B_m$ additionally preserving a direct path from their observed features.

\subsection{Global Context Modulation}\label{sec:msrgf_supp_gcm}

GCM converts the enhanced sequences from RGF into a sample-level prediction by combining modality summaries with a shared global context. It first pools the valid positions of each $\widetilde H_m$, then uses the cross-modality context to adjust the modality-specific summaries before prediction. Specifically, each modality has a learned pooling query $q_m^{\mathrm{pool}}$ that reads the valid positions of $\widetilde H_m$ through four-head attention. For a modality $m$, the pooling operation is:
\begin{equation}
\label{eq:msrgf_supp_gcm_pool}
\begin{aligned}
z_m&=\operatorname{MHA}_m\big(q_m^{\mathrm{pool}},
\operatorname{LN}_m(\widetilde H_m),
\operatorname{LN}_m(\widetilde H_m);P_m\big),\\
v_m&=\Pi_m(W_m^{\mathrm{pool}}z_m+b_m^{\mathrm{pool}}).
\end{aligned}
\end{equation}

Both $W_m^{\mathrm{pool}}z_m+b_m^{\mathrm{pool}}$ and $\Pi_m$ are modality-specific linear projections from 128 to 128 dimensions, with distinct parameters; $\Pi_m$ maps the pooling output to the common fusion space. A separate four-head attention layer reads the three summaries $[v_T,v_A,v_V]$ through the learned query $q_c$ to produce the global context $c$; the following MLP $f_c$ has widths $128\!\to\!128\!\to\!128$ with GELU.

For each modality, $s_m=W_m^{\mathrm{back}}c$ projects the global context to the same width as $v_m$. The shared MLP $f_\gamma$ uses $[v_m,s_m,v_m-s_m,v_m\odot s_m]$ together with a learned modality embedding $e_m^\gamma\in\mathbb R^{512}$ to compute $\gamma_m$. This shared network has widths $512\!\to\!128\!\to\!128$ followed by sigmoid, with its final bias initialized to $-2$. The prediction MLP $f_{\mathrm{pred}}$ receives $[c,u_T,u_A,u_V]$ and uses linear layers of widths $512\!\to\!256\!\to\!128\!\to\!d_y$, with GELU and dropout $0.1$. Here $d_y=1$ for the two sentiment datasets and $d_y=8$ for the four two-logit emotion outputs of IEMOCAP.

\subsection{Network Configuration and Parameter Sharing}\label{sec:msrgf_supp_arch_summary}
Table~\ref{tab:msrgf_supp_arch_summary} consolidates the architectural configurations introduced above, including the dimensions of the attention and feed-forward components and the parameters shared across source--target paths. All hidden sequences use $d=128$, with modality-specific projections and output mappings retained where the input and output spaces differ.
\begin{table*}[t]
\centering
\caption{Network configuration and parameter sharing in MSRGF.}
\label{tab:msrgf_supp_arch_summary}
\small
\setlength{\tabcolsep}{3pt}
\renewcommand{\arraystretch}{1.10}
\begin{tabular*}{\textwidth}{@{\extracolsep{\fill}}p{0.09\textwidth}p{0.29\textwidth}p{0.55\textwidth}@{}}
\toprule
Module & Network & Configuration and parameter sharing \\
\midrule
Input & Projection and temporal encoder & $d_m\to128$ + LayerNorm; one 4-head Transformer per modality \\
SWR & Prompt bank and query network & Four $128$-dimensional prompts per modality; shared $128\to128\to128$ MLP \\
SWR & Reconstruction generator & Two 4-head cross-attention blocks, each with $128\to256\to128$ FFN; shared across paths \\
SWR & Output MLPs and decoders & Three modality-specific $128\to128\to128$ GELU MLPs; three modality-specific linear $128\to d_m$ decoders (with bias), reused in RGF \\
RGF & Base representation & One shared cross-attention block and $128\to128\to128$ MLP \\
RGF & Relation comparison & Three modality-specific linear $128\to64$ projections; shared $256\to64\to16$ GELU interaction MLP; binary availability indicators \\
RGF & Reference and refinement weights & Two separate shared $279\to128\to1$ scoring MLPs \\
RGF & Feature gate and difference network & Shared $407\to128\to128$ gate; bias-free $128\to256\to128$ MLP \\
RGF & Integration & Shared $408\to128\to256$ MLP; $\eta_a=\eta_g=0.1$ \\
GCM & Modality and global attention & Three modality-specific 4-head poolers and global 4-head attention \\
GCM & Modality projections & Per modality: distinct $W_m^{\mathrm{pool}}$ and $\Pi_m$ ($128\to128$), and bias-free $W_m^{\mathrm{back}}$ ($128\to128$) \\
GCM & Context, gate, and prediction MLPs & $f_c:128\to128\to128$ GELU; shared $f_\gamma:512\to128\to128$ GELU + sigmoid; $f_{\mathrm{pred}}:512\to256\to128\to d_y$ GELU + dropout \\
\bottomrule
\end{tabular*}
\end{table*}

\section{Training Details}\label{sec:msrgf_supp_training}
The training objective couples task supervision with losses on the individual SWR estimates and the enhanced representations produced by RGF. Because these quantities span modalities with different feature widths and observation counts, we first describe how auxiliary losses are normalized, then trace their gradients through the network and specify the optimization schedule.

\subsection{Loss Normalization}
The view, reconstruction, and preservation objectives operate on features with different dimensions and on sets of observed or missing positions of different sizes. We normalize each error over its feature dimension and eligible temporal positions before averaging over the contributing sample--modality pairs. For sample $i$, modality $m$, and a position mask $W_{im}$, we first average absolute differences over the feature dimension $d_U$ and then over the selected temporal positions:
\begin{equation}
\label{eq:msrgf_supp_masked_l1}
\ell_{im}(U,V;W)=
\frac{\sum_tW_{im}(t)\,\|U_{im}(t)-V_{im}(t)\|_1/d_U}
{\max(\sum_tW_{im}(t),1)}.
\end{equation}

Let $\mathcal E_\Omega=\{(i,m):\|\Omega_{im}\|_1>0\}$ and $\mathcal E_M=\{(i,m):\|M_{im}\|_1>0\}$ denote the sample--modality pairs that contribute to reconstruction and preservation, respectively. The two reconstruction losses follow the representations produced by SWR and RGF:
\begin{equation}
\label{eq:msrgf_supp_view_loss}
\begin{aligned}
\mathcal L_{\mathrm{view}}={}&
\frac1{|\mathcal E_\Omega|}
\sum_{(i,m)\in\mathcal E_\Omega}
\frac1{|\mathcal O_i|}\\
&\cdot\sum_{n\in\mathcal O_i}
\ell_{im}\big(D_m(R_{im}^n),X_{im}^\star;\Omega_{im}\big).
\end{aligned}
\end{equation}
\begin{equation}
\label{eq:msrgf_supp_rec_loss}
\mathcal L_{\mathrm{rec}}=
\frac1{|\mathcal E_\Omega|}
\sum_{(i,m)\in\mathcal E_\Omega}
\ell_{im}\big(D_m(\widetilde H_{im}),X_{im}^\star;\Omega_{im}\big).
\end{equation}

For $\mathcal L_{\mathrm{view}}$, the errors from available modalities are averaged within each sample--modality pair before the final mean. The preservation loss uses the positions observed by the model:
\begin{equation}
\label{eq:msrgf_supp_pres_loss}
\mathcal L_{\mathrm{pres}}=
\frac1{|\mathcal E_M|}
\sum_{(i,m)\in\mathcal E_M}
\ell_{im}\big(\widetilde H_{im},Z_{im};M_{im}\big).
\end{equation}
When either eligible set is empty, the corresponding loss is set to zero. The projected features $Z_m$ remain trainable in the preservation loss. The sentiment loss and the total objective follow Section~\ref{sec:msrgf_losses} of the main paper.

\subsection{Sentiment Supervision and Gradient Flow}\label{sec:msrgf_supp_gradients}
While the auxiliary objectives supervise reconstruction and preservation at different points in the pipeline, the task loss trains their resulting representations for sentiment or emotion prediction. On CMU-MOSI and CMU-MOSEI, the model uses $L_1$ regression over the original sentiment scores. IEMOCAP uses the average cross-entropy of four binary emotion predictions, with two logits for each emotion. These objectives supervise the final prediction from GCM, and their gradients pass through the enhanced representations in RGF and the estimates in SWR to the active input encoders.

The auxiliary losses supervise different points along the same computation. $\mathcal L_{\mathrm{view}}$ updates the available reconstruction paths and the decoder $D_m$ through their individual estimates $R_m^n$. $\mathcal L_{\mathrm{rec}}$ supervises the enhanced representations $\widetilde H_m$, passing gradients through the RGF weights and gates, the SWR estimates, and the shared decoder. $\mathcal L_{\mathrm{pres}}$ acts at observed positions on $\widetilde H_m$ and the trainable projected features $Z_m$. The combined objective therefore connects source-wise estimation, final representation construction, and sentiment prediction.

Within the relation descriptor, gradients pass through the learned source--estimate interaction network. The cosine scores measuring observed-feature agreement and differences among pooled estimates use the stop-gradient operation $\operatorname{sg}(\cdot)$. Their values still enter the subsequent scoring networks, which remain trainable alongside the difference transform and integration gates. The decoder also provides the features used by the observed-feature agreement score and receives direct gradients from the reconstruction objectives.

\subsection{Implementation Details}\label{sec:implementation_details}
The input features and missingness protocols are detailed in Section~\ref{sec:msrgf_supp_inputs}, the SWR, RGF, and GCM architectures in Sections~\ref{sec:msrgf_supp_swr}--\ref{sec:msrgf_supp_gcm}, and Table~\ref{tab:msrgf_supp_arch_summary}. Here, we provide the remaining training and evaluation settings.

\textbf{Optimization.} We use AdamW with $(\beta_1,\beta_2)=(0.9,0.999)$ and $\epsilon=10^{-8}$. Weight decay is 0.01 for eligible matrix parameters and zero for biases and normalization parameters. The default learning rate for MSRGF is $10^{-4}$; the default learning rates for the text encoder are $6\times10^{-6}$ on CMU-MOSI and $3\times10^{-6}$ on CMU-MOSEI. The default batch size is 32. Each optimization phase uses 10\% warm-up followed by linear learning-rate decay to a minimum multiplier of 0.1, with global gradient-norm clipping at 1.0. The default auxiliary-loss coefficients in Equation~\ref{eq:msrgf_total} of the main paper are $(\lambda_{rec},\lambda_{view},\lambda_{pres})=(0.1,0.1,0.05)$.

\textbf{Training schedule.} On CMU-MOSI and CMU-MOSEI, the text encoder is frozen for two adaptation epochs and then jointly optimized with MSRGF for up to 40 epochs. A new optimizer is initialized at the stage transition. The encoder is DeBERTa-v3-large, and word-level features are obtained by averaging its subword representations. During joint training, the encoder remains in evaluation mode to disable dropout while preserving gradients.

\textbf{Evaluation.} For CMU-MOSI and CMU-MOSEI, binary Acc and weighted F1 exclude zero-valued sentiment labels as in prior works \cite{zhuang2026hyper,zhuang2025cmad}, which remain in regression training; for IEMOCAP, Acc and weighted F1 are averaged over the four binary emotion tasks as in \cite{zhuang2026hyper,zhuang2025cmad}.

\section{Additional Experimental Results}\label{sec:msrgf_supp_results}
This section expands the comparisons by reporting Acc/F1 at every evaluated random missing rate. Random whole-modality missingness changes the available sources from which SWR can construct estimates, whereas random within-modality missingness changes the temporal observations supplied to the source encoders. Results are presented separately for the two protocols, with Avg calculated as the arithmetic mean across the rates shown; the following section then examines the model components responsible for these predictions.

\subsection{Random Whole-Modality Missingness}\label{sec:msrgf_supp_whole}
Table~\ref{tab:MSRGF_whole_all} evaluates random whole-modality missingness on CMU-MOSEI and IEMOCAP, where one or more modalities may be entirely unavailable but each sample retains at least one source. The seven reported rates cover 10\%--70\% missingness, with the highest setting capped at the attainable rate for three modalities.
\begin{table*}[t]
\centering
\caption{Random whole-modality missingness (Acc/F1, \%). The 70\% setting is the approximation of 2/3. Best and second-best are bold and underlined; $^{\ast}$ denotes results from HME~\cite{zhuang2026hyper}.}
\label{tab:MSRGF_whole_all}
\small
\setlength{\tabcolsep}{2.0pt}
\renewcommand{\arraystretch}{1.10}
\begin{tabular*}{\textwidth}{@{\extracolsep{\fill}}lcccccccc@{}}
\toprule
Method & 10\% & 20\% & 30\% & 40\% & 50\% & 60\% & 70\% & Avg \\
\midrule
\multicolumn{9}{@{}l}{\textbf{CMU-MOSEI}} \\
DiCMoR$^{\ast}$~\cite{wang2023dicmor} & 83.2/83.1 & 81.2/80.8 & 79.0/77.9 & 76.9/75.2 & 73.7/71.7 & 71.1/70.4 & 70.6/68.9 & 76.5/75.4 \\
IMDer$^{\ast}$~\cite{wang2023imder} & 83.5/83.3 & 81.3/80.9 & 79.4/78.8 & 76.1/74.3 & 75.3/72.4 & 71.2/66.5 & 71.1/65.9 & 76.8/74.6 \\
LNLN$^{\ast}$~\cite{zhang2024lnln} & 83.1/83.0 & 81.7/81.0 & 79.1/78.1 & 76.9/75.6 & 75.1/72.4 & 72.5/68.6 & 70.6/65.2 & 77.0/74.8 \\
MPLMM$^{\ast}$~\cite{guo2024mplmm} & 83.7/83.5 & 81.4/80.8 & 79.2/78.3 & 77.1/76.1 & 75.6/73.8 & 72.7/70.6 & 70.5/69.2 & 77.2/76.0 \\
HME$^{\ast}$~\cite{zhuang2026hyper} & 84.0/83.8 & 82.3/81.9 & 80.0/79.6 & 77.6/76.4 & \underline{75.7}/74.7 & 73.0/72.5 & \underline{72.4}/69.2 & 77.9/76.9 \\
API~\cite{wang2026api} & \underline{84.1}/\underline{84.1} & \underline{82.9}/\underline{82.8} & \underline{80.9}/\underline{81.1} & \underline{78.5}/\underline{78.3} & 75.6/\underline{75.3} & \underline{74.2}/\textbf{73.7} & \underline{72.4}/\underline{69.3} & \underline{78.4}/\underline{77.8} \\
\midrule
\textbf{MSRGF} & \textbf{84.9}/\textbf{84.7} & \textbf{83.2}/\textbf{83.0} & \textbf{81.6}/\textbf{81.3} & \textbf{79.2}/\textbf{78.5} & \textbf{76.6}/\textbf{75.9} & \textbf{74.7}/\underline{73.5} & \textbf{72.7}/\textbf{71.0} & \textbf{79.0}/\textbf{78.3} \\
\midrule
\multicolumn{9}{@{}l}{\textbf{IEMOCAP}} \\
DiCMoR$^{\ast}$~\cite{wang2023dicmor} & 77.1/73.7 & 76.4/72.5 & 75.8/71.6 & 75.5/69.0 & 75.1/68.5 & 74.5/68.0 & 74.3/67.5 & 75.5/70.1 \\
IMDer$^{\ast}$~\cite{wang2023imder} & 77.2/72.8 & 75.9/71.6 & 75.7/70.8 & 74.9/70.8 & 74.7/68.2 & 73.5/67.0 & 73.1/65.3 & 75.0/69.5 \\
LNLN$^{\ast}$~\cite{zhang2024lnln} & 78.0/76.3 & 77.6/75.7 & 77.5/75.4 & 77.4/74.0 & 76.8/72.9 & 76.6/72.6 & 76.5/70.2 & 77.2/73.9 \\
MPLMM$^{\ast}$~\cite{guo2024mplmm} & 76.1/70.8 & 75.7/67.9 & 75.6/67.4 & 75.5/66.6 & 74.0/66.2 & 73.2/66.0 & 72.4/65.7 & 74.6/67.2 \\
HME$^{\ast}$~\cite{zhuang2026hyper} & \underline{80.9}/\underline{80.3} & \underline{80.7}/\underline{80.0} & \underline{79.0}/\underline{77.8} & \underline{78.7}/\underline{77.3} & \underline{78.5}/\underline{77.1} & \underline{77.8}/\textbf{76.1} & \underline{77.6}/\underline{76.0} & \underline{79.0}/\underline{77.8} \\
\midrule
\textbf{MSRGF} & \textbf{81.2}/\textbf{80.5} & \textbf{80.9}/\textbf{80.2} & \textbf{80.0}/\textbf{79.4} & \textbf{79.1}/\textbf{77.7} & \textbf{79.5}/\textbf{77.6} & \textbf{78.2}/\underline{75.5} & \textbf{78.5}/\textbf{76.8} & \textbf{79.6}/\textbf{78.2} \\
\bottomrule
\end{tabular*}
\end{table*}

On CMU-MOSEI, MSRGF achieves average Acc/F1 of 79.0/78.3, compared with 78.4/77.8 for API~\cite{wang2026api}, and leads in accuracy at all seven rates. Its accuracy remains at 76.6\% for 50\% missingness and 72.7\% at the capped 70\% setting. On IEMOCAP, MSRGF reaches 79.6/78.2 on average, improving over HME~\cite{zhuang2026hyper} by 0.6/0.4 points.

\subsection{Random Within-Modality Missingness}\label{sec:msrgf_supp_within}
To test incomplete observations within otherwise available modalities, Table~\ref{tab:MSRGF_intra_all} reports the corresponding Acc/F1 results on CMU-MOSI and CMU-MOSEI for rates from 10\% to 90\%. Unlike whole-modality removal, this protocol retains scattered temporal observations that remain accessible to SWR and the observation-conditioned fusion in RGF.
\begin{table*}[t]
\centering
\caption{Random within-modality missingness (Acc/F1, \%). Best and second-best results are bold and underlined.}
\label{tab:MSRGF_intra_all}
\scriptsize
\setlength{\tabcolsep}{0.85pt}
\renewcommand{\arraystretch}{1.10}
\begin{tabular*}{\textwidth}{@{\extracolsep{\fill}}lcccccccccc@{}}
\toprule
Method & 10\% & 20\% & 30\% & 40\% & 50\% & 60\% & 70\% & 80\% & 90\% & Avg \\
\midrule
\multicolumn{11}{@{}l}{\textbf{CMU-MOSI}} \\
ALMT~\cite{zhang2023almt} & 80.9/81.3 & 77.6/77.9 & 75.2/75.5 & 73.1/73.9 & 68.2/69.7 & 64.5/66.8 & 61.8/65.3 & 60.4/\textbf{65.5} & 57.3/\textbf{64.9} & 68.8/71.2 \\
LNLN~\cite{zhang2024lnln} & 81.2/81.6 & 79.2/79.5 & 77.3/77.6 & 76.0/76.3 & 73.4/73.7 & \underline{69.0}/69.2 & 66.0/66.0 & 62.8/62.6 & 56.5/56.3 & 71.3/71.4 \\
P-RMF~\cite{zhu2025prmf} & 82.6/82.9 & 79.6/81.0 & 76.8/\underline{79.3} & 75.5/\underline{77.7} & 73.0/73.3 & 68.8/68.6 & 66.2/64.7 & 62.0/60.8 & 59.5/56.7 & 71.6/71.7 \\
DEAR~\cite{zou2026dear} & 80.6/82.5 & 78.6/80.4 & 77.5/78.4 & 74.8/76.1 & 72.9/74.1 & 68.0/\textbf{69.8} & 64.9/\underline{67.1} & 62.4/63.4 & 59.1/60.4 & 71.0/\underline{72.5} \\
SemMSA~\cite{li2026semmsa} & \underline{84.9}/\underline{84.7} & \underline{81.6}/\underline{81.4} & \underline{78.8}/78.7 & \textbf{78.0}/\textbf{77.9} & \textbf{75.6}/\textbf{75.5} & \underline{69.0}/69.2 & 64.8/65.0 & 62.8/63.0 & 56.0/55.2 & \underline{72.4}/72.3 \\
TF-Mamba~\cite{li2025tfmamba} & 81.9/81.9 & 80.5/80.6 & 77.7/77.9 & 76.1/76.2 & \underline{75.0}/\underline{75.2} & 68.6/68.8 & \underline{67.2}/\textbf{67.4} & \underline{63.6}/63.8 & \underline{60.4}/60.6 & 72.3/\underline{72.5} \\
\midrule
\textbf{MSRGF} & \textbf{87.8}/\textbf{87.8} & \textbf{82.2}/\textbf{82.2} & \textbf{81.6}/\textbf{81.6} & \underline{76.4}/76.4 & 73.8/73.9 & \textbf{69.2}/\underline{69.4} & \textbf{67.4}/\textbf{67.4} & \textbf{64.2}/\underline{64.3} & \textbf{64.2}/\underline{64.2} & \textbf{74.1}/\textbf{74.1} \\
\midrule
\multicolumn{11}{@{}l}{\textbf{CMU-MOSEI}} \\
ALMT~\cite{zhang2023almt} & 84.1/84.2 & 82.7/82.8 & 80.9/81.2 & 79.4/79.7 & 77.4/77.7 & 75.0/75.4 & 71.6/72.2 & 68.2/69.1 & 61.4/63.3 & 75.6/76.2 \\
LNLN~\cite{zhang2024lnln} & 83.3/83.7 & 81.7/82.0 & 80.1/80.4 & 78.5/79.0 & 76.4/77.2 & 73.8/75.0 & 71.6/73.5 & 68.6/71.8 & 64.8/\underline{70.6} & 75.4/77.0 \\
P-RMF~\cite{zhu2025prmf} & 84.0/84.4 & 83.0/83.4 & 81.3/81.9 & 80.0/80.7 & 78.6/79.5 & 76.1/77.6 & 74.6/75.9 & 70.8/\underline{73.0} & 67.9/\textbf{71.5} & 77.4/78.7 \\
DEAR~\cite{zou2026dear} & 84.3/84.9 & 83.2/83.3 & 82.3/\underline{82.9} & 81.4/80.9 & 79.6/79.8 & 77.9/76.8 & \underline{75.5}/74.7 & \underline{73.2}/71.8 & \underline{70.0}/69.6 & \underline{78.6}/78.3 \\
SemMSA~\cite{li2026semmsa} & \underline{85.3}/\underline{85.2} & \underline{84.0}/\underline{83.9} & \underline{82.6}/82.4 & \underline{81.6}/\underline{81.4} & \underline{80.9}/\underline{80.6} & \underline{79.0}/\underline{78.6} & \underline{75.5}/\underline{76.8} & 70.4/\underline{73.0} & 68.2/\underline{70.6} & \underline{78.6}/\underline{79.2} \\
TF-Mamba~\cite{li2025tfmamba} & 83.2/83.0 & 82.6/82.4 & 80.8/80.6 & 80.0/79.8 & 78.6/78.3 & 75.9/75.6 & 73.5/73.3 & 70.3/70.2 & 64.8/64.9 & 76.6/76.5 \\
\midrule
\textbf{MSRGF} & \textbf{86.9}/\textbf{86.9} & \textbf{86.0}/\textbf{86.0} & \textbf{85.0}/\textbf{84.8} & \textbf{84.2}/\textbf{84.1} & \textbf{81.1}/\textbf{81.1} & \textbf{79.6}/\textbf{79.1} & \textbf{77.7}/\textbf{77.6} & \textbf{74.3}/\textbf{73.6} & \textbf{70.1}/69.0 & \textbf{80.5}/\textbf{80.2} \\
\bottomrule
\end{tabular*}
\end{table*}

Across the nine within-modality rates, MSRGF averages 74.1/74.1 Acc/F1 on CMU-MOSI and 80.5/80.2 on CMU-MOSEI, exceeding the strongest baseline average accuracy by 1.7 and 1.9 points, respectively. The improvements extend to high missingness, where accuracy reaches 64.2\% on MOSI and 70.1\% on MOSEI at 90\% masking. Because some observed positions remain in each modality, this experiment complements whole-modality removal by evaluating how the source encoders and reference-guided integration respond to increasingly sparse temporal input.

\section{Additional Ablation Studies and Mechanism Analysis}\label{sec:msrgf_supp_mech}
The performance comparisons above establish the behavior of MSRGF across missingness protocols; the remaining question is how its individual computations contribute to these outcomes. 

\subsection{Retrained Design Ablations}\label{sec:msrgf_supp_retrained}
We first examine whether SWR generates useful source-conditioned estimates, whether RGF improves their integration, and whether GCM contributes to sentiment prediction. Each modified architecture or objective is independently retrained on IEMOCAP, with Acc/F1 averaged over seven fixed modality combinations and seven random whole-modality missing rates, respectively. Beyond the major components, Table~\ref{tab:msrgf_ablation_add} evaluates the internal designs of reference construction, relation modeling, difference refinement, observed-feature integration, context adjustment, and auxiliary objectives.

\begin{table}[t]
\centering
\caption{Retrained IEMOCAP ablations (Acc/F1, \%); Fixed and Random average seven settings each.}
\label{tab:msrgf_ablation_add}
\small
\setlength{\tabcolsep}{1.2pt}
\renewcommand{\arraystretch}{1.10}
\begin{tabular*}{\columnwidth}{@{\extracolsep{\fill}}p{0.47\columnwidth}cc@{}}
\toprule
Method & Fixed & Random \\
\midrule
\multicolumn{3}{@{}l}{\textbf{Major Components}} \\
\midrule
w/o SWR & 78.09/75.98 & 77.64/75.17 \\
w/o RGF & 78.05/75.81 & 77.74/75.27 \\
w/o GCM & 78.25/76.00 & 77.94/75.63 \\
\midrule
\multicolumn{3}{@{}l}{\textbf{Source Organization}} \\
\midrule
T-priority Single Source
    & 78.38/76.09 & 78.08/75.82 \\
A-priority Single Source
    & 78.25/75.85 & 78.03/75.73 \\
V-priority Single Source
    & 78.37/76.19 & 77.86/75.55 \\
Fused-source Generator
    & 78.23/76.12 & 78.05/75.72 \\
Early Aggregation
    & 78.43/76.16 & 78.05/75.74 \\
\midrule
\multicolumn{3}{@{}l}{\textbf{Internal Mechanisms}} \\
\midrule
w/o Target Queries & 78.27/76.14 & 77.96/75.75 \\
w/o Contextual Anchor & 78.01/75.79 & 77.78/75.37 \\
Uniform $\alpha$ (reference only) & 78.38/76.12 & 78.05/75.74 \\
Uniform $\beta$ & 78.29/76.02 & 78.03/75.72 \\
$\beta=\alpha$ & 78.35/76.07 & 78.06/75.70 \\
w/o Difference Refinement & 78.34/76.06 & 78.05/75.72 \\
Ordinary Residual & 78.30/76.02 & 78.03/75.75 \\
w/o Relation Descriptor & 78.46/76.16 & 78.08/75.81 \\
w/o $c^{\mathrm{src}}$ & 78.68/77.15 & 78.88/77.49 \\
w/o $c^{\mathrm{div}}$ & 79.02/77.52 & 78.91/77.51 \\
w/o $c^{\mathrm{tar}}$ & 78.96/77.47 & 78.96/77.57 \\
w/o Feature-wise Gates & 78.27/76.00 & 78.02/75.72 \\
Hard Observation Copy & 77.96/75.76 & 77.83/75.38 \\
w/o Observed Fusion & 78.40/76.79 & 78.13/76.35 \\
w/o Context Gates & 78.34/76.10 & 78.08/75.85 \\
\midrule
\multicolumn{3}{@{}l}{\textbf{Loss Functions}} \\
\midrule
w/o $\mathcal L_{\mathrm{view}}$ & 78.20/75.99 & 78.07/75.79 \\
w/o $\mathcal L_{\mathrm{rec}}$ & 78.20/75.99 & 78.08/75.81 \\
w/o $\mathcal L_{\mathrm{pres}}$ & 78.38/76.44 & 78.22/75.98 \\
\midrule
\textbf{MSRGF} & \textbf{79.73/78.17} & \textbf{79.63/78.25} \\
\bottomrule
\end{tabular*}
\end{table}

\textbf{Major components and target representations.} Removing SWR bypasses the reconstructed views and uses $\widetilde H_m=B_m$, with $\mathcal L_{\mathrm{view}}$ and $\mathcal L_{\mathrm{rec}}$ disabled. Replacing RGF with uniform completion retains $Z_m$ at observed positions and averages the available source estimates at missing positions. Removing GCM preserves modality pooling and the prediction network but sets $c=\mathbf{0}$ and $u_m=v_m$. Under Random, these variants achieve F1 scores of 75.17, 75.27, and 75.63, respectively, compared with 78.25 for MSRGF, with similar reductions under Fixed. We further examine the target representations by removing the target queries ($Q_m=\mathbf{0}$) or replacing the contextual base with $B_m=Z_m$. Both modifications reduce performance, indicating that target-specific queries and the contextual base contribute to subsequent reconstruction and fusion.

\textbf{Reconstruction source organization.} We further examine whether separate source-conditioned estimates are beneficial compared with single-source reconstruction or joint source processing. Three priority single-source variants select the first available modality according to fixed preference orders $\pi_T=(T,A,V)$, $\pi_A=(A,T,V)$, and $\pi_V=(V,T,A)$. The selected source provides the only reconstruction estimate for every target modality, including observed targets, while the original observed inputs and their direct integration pathway are retained. With one active estimate, the reference equals that estimate and difference refinement is inactive. We also evaluate Fused-source Generator, which concatenates the encoded modality features with source embeddings and applies masked joint attention to produce a single target estimate. Unlike Early Aggregation, which averages separately generated estimates, this variant combines source information within the generator. All variants are independently retrained, retaining the applicable reconstruction objectives and downstream fusion and prediction components. As shown in Table~\ref{tab:msrgf_ablation_add}, the priority single-source variants achieve F1 scores of 75.85--76.19 under Fixed and 75.55--75.82 under Random, compared with 78.17 and 78.25 for MSRGF. Fused-source Generator obtains F1 scores of 76.12 and 75.72, close to Early Aggregation at 76.16 and 75.74. These results indicate that retaining separate source-conditioned estimates for subsequent reference-guided fusion is more effective than the evaluated single-source and early-fusion alternatives.

\textbf{Reference construction and difference refinement.} We next examine the weighting and refinement mechanisms within RGF. Uniform $\alpha$ (reference only) replaces the learned reference weights with uniform coefficients when computing $C_m$, while Uniform $\beta$ assigns equal refinement weights and $\beta=\alpha$ ties them to the reference weights. To examine the source-specific update, w/o Difference Refinement sets $N_m=\mathbf{0}$, Ordinary Residual replaces $R_m^n-C_m$ with the shared residual $C_m-B_m$, and Early Aggregation averages the estimates before refinement, eliminating their source-specific differences. All alternatives yield lower F1 than MSRGF under Fixed and Random. These results indicate that learned reference weighting and source-specific difference refinement improve integration compared with the evaluated alternatives.

\textbf{Relation descriptor and its components.} We further investigate the relation descriptor $\xi_m^n$ and its individual comparison features. The w/o Relation Descriptor variant sets $\xi_m^n=\mathbf{0}$, while three additional variants independently remove the 16-dimensional source--estimate interaction $c_{m,n}^{\mathrm{src}}$, the cross-estimate difference score $c_{m,n}^{\mathrm{div}}$, and the observed-target agreement score $c_{m,n}^{\mathrm{tar}}$. Each feature is set to zero while retaining the remaining descriptor entries, validity indicators, and subsequent fusion operations.

Removing the entire descriptor decreases F1 by 2.01/2.44 points under Fixed/Random, while removing its individual features produces smaller but consistent reductions. Specifically, w/o $c^{\mathrm{src}}$ reduces F1 by 1.02/0.76 points, indicating that source--estimate interactions contribute to fusion. Removing $c^{\mathrm{div}}$ decreases F1 by 0.65/0.74 points, although the source-specific differences $\Delta_m^n$ and their refinement $N_m$ remain active. This indicates that explicitly measured differences among source estimates provide additional information for reference construction and refinement. Similarly, removing $c^{\mathrm{tar}}$ reduces F1 by 0.70/0.68 points, showing the contribution of observed-target agreement when such comparisons are available. Together, these results indicate that the relation descriptor incorporates complementary comparison features that improve the learned weighting and gating.

\textbf{Observed-feature integration.} We then examine how RGF incorporates retained observations. Hard Observation Copy forces $\widetilde H_m(t)=Z_m(t)$ at observed positions after integration, whereas w/o Observed Fusion uses $\widetilde H_m=C_m+g_m^{\mathrm{enh}}\odot N_m$, retaining $B_m$ only for weighting and gating. Both variants reduce performance, with Random F1 decreasing to 75.38 and 76.35, respectively. These results indicate that integrating observed and reconstructed information through the learned fusion mechanism improves prediction compared with strict copying or removing the direct observation pathway. Hard copying also makes the preservation loss zero at the copied positions, so this ablation jointly changes the representation and its training constraint.

\textbf{Context adjustment.} Beyond removing the complete GCM pathway, we examine its learned modulation by fixing $\gamma_m=\mathbf{1}$ in w/o Context Gates while retaining global context construction and the prediction pathway. This modification reduces F1 to 76.10/75.85 under Fixed/Random, compared with 78.17/78.25 for MSRGF. The results indicate that the learned context gates contribute to prediction in addition to the global context representation.

\textbf{Auxiliary objectives.} Finally, we examine the three auxiliary objectives while retaining their associated model computations. Removing $\mathcal L_{\mathrm{view}}$ eliminates direct supervision of the source-wise estimates $R_m^n$, removing $\mathcal L_{\mathrm{rec}}$ disables reconstruction supervision on the integrated representation $\widetilde H_m$, and removing $\mathcal L_{\mathrm{pres}}$ eliminates the constraint at observed positions. All three removals reduce performance under Fixed and Random, with Random F1 reaching 75.79, 75.81, and 75.98, respectively. These results complement the architectural ablations by examining the training objectives used to learn the reconstructed and enhanced representations.

\subsection{Contribution of Source-Wise Estimates}\label{sec:msrgf_supp_sources}
We further examine how individual source estimates contribute within the trained MSRGF model. The component ablation identifies the importance of SWR, but it does not show how different source estimates are used after training. We therefore restrict inference to one source estimate for an entirely missing target while leaving the original observations unchanged. As shown in Table~\ref{tab:msrgf_supp_source_task}, combining the two estimates increases F1 from the better single-source value of 79.20 to 80.61 under TV and from 77.03 to 77.58 under AV. The comparison isolates the effect of retaining multiple estimates in the trained fusion pipeline without removing the observed modalities.

\begin{table}[t]
\centering
\caption{IEMOCAP task results using one or both source estimates (Acc/F1, \%).}
\label{tab:msrgf_supp_source_task}
\small
\setlength{\tabcolsep}{3pt}
\renewcommand{\arraystretch}{1.10}
\begin{tabular*}{\columnwidth}{@{\extracolsep{\fill}}llcc@{}}
\toprule
Setting & Source estimate & One estimate & Both estimates \\
\midrule
TV & $R_A^T$ & 79.51/79.20 & 80.85/80.61 \\
TV & $R_A^V$ & 79.18/78.70 & 80.85/80.61 \\
AV & $R_T^A$ & 79.21/77.03 & 79.73/77.58 \\
AV & $R_T^V$ & 77.32/73.80 & 79.73/77.58 \\
\bottomrule
\end{tabular*}
\end{table}

To examine the representation quality behind these task changes, we compare decoded features with the complete target features using the modality-specific decoder $D_m$. Table~\ref{tab:msrgf_supp_single_source_reconstruction} shows that the enhanced visual representation has lower MAE and higher cosine similarity than the observation base when text or audio is the sole source. With multiple available sources, Table~\ref{tab:msrgf_supp_reconstruction} further distinguishes the individual estimates from their reference and enhanced representations. For the missing textual and acoustic targets, the learned reference yields the lowest decoded MSE among the displayed alternatives, although the final enhanced representation is optimized jointly for reconstruction and the task objective and need not minimize each reconstruction metric. All reconstruction metrics are compared within the same target modality.

\begin{table}[t]
\centering
\caption{Decoded visual-target reconstruction on IEMOCAP.}
\label{tab:msrgf_supp_single_source_reconstruction}
\footnotesize
\setlength{\tabcolsep}{3pt}
\renewcommand{\arraystretch}{1.08}
\begin{tabular*}{\columnwidth}{@{\extracolsep{\fill}}llccc@{}}
\toprule
Source & Representation & MAE & MSE & Cosine \\
\midrule
T$\rightarrow$V & $B_V$ & 0.583169 & 0.674870 & 0.060257 \\ & $\widetilde H_V$ & \textbf{0.365323} & \textbf{0.674498} & \textbf{0.621105} \\
\addlinespace[2pt]
A$\rightarrow$V & $B_V$ & 0.584370 & 0.673633 & $-0.017629$ \\ & $\widetilde H_V$ & \textbf{0.448677} & \textbf{0.635592} & \textbf{0.547529} \\
\bottomrule
\end{tabular*}
\end{table}

\begin{table}[t]
\centering
\caption{Decoded reconstruction of missing IEMOCAP targets; bold marks the best within each target.}
\label{tab:msrgf_supp_reconstruction}
\small
\setlength{\tabcolsep}{7pt}
\renewcommand{\arraystretch}{1.08}
\begin{tabular}{@{}llccc@{}}
\toprule
Ava $\rightarrow$ Tar & Repr & MAE $\downarrow$ & MSE $\downarrow$ & Cosine $\uparrow$ \\
\midrule
AV$\rightarrow$T & $B_T$ & 0.779919 & 1.039006 & $-0.007759$ \\ & $R_T^A$ & 0.770692 & 1.029792 & 0.054224 \\ & $R_T^V$ & 0.771048 & 1.030405 & 0.054383 \\ & $C_T$ & \textbf{0.770538} & \textbf{1.029438} & \textbf{0.055783} \\ & $\widetilde H_T$ & 0.770644 & 1.029687 & 0.055194 \\
\addlinespace[3pt]
TV$\rightarrow$A & $B_A$ & 0.676690 & 0.874101 & 0.011633 \\ & $R_A^T$ & 0.645782 & 0.871201 & \textbf{0.123966} \\ & $R_A^V$ & 0.666480 & 0.930592 & 0.042970 \\ & $C_A$ & \textbf{0.643271} & \textbf{0.867499} & 0.118265 \\ & $\widetilde H_A$ & 0.648512 & 0.881017 & 0.117184 \\
\bottomrule
\end{tabular}
\end{table}

\begin{table*}[t]
\centering
\caption{IEMOCAP F1 (\%) after interventions on $U_m=g_m^{\mathrm{enh}}\odot N_m$.}
\label{tab:msrgf_supp_refinement_task}
\small
\setlength{\tabcolsep}{4pt}
\renewcommand{\arraystretch}{1.10}
\begin{tabular*}{\textwidth}{@{\extracolsep{\fill}}lccccc@{}}
\toprule
Target & Full & $U_m=0$ & Random & Reversed & Permuted \\
\midrule
AV$\rightarrow$T & \textbf{77.58} & 77.33 & 77.28 & 76.85 & 76.88 \\
TV$\rightarrow$A & \textbf{80.61} & 80.00 & 80.03 & 79.29 & 79.60 \\
\bottomrule
\end{tabular*}
\end{table*}

\subsection{Reference and Relation Interventions}\label{sec:msrgf_supp_relation}
Having examined the estimates supplied by SWR, we next consider the relation information used to weight them in RGF. The retrained descriptor ablation tests the complete feature $\xi_m^n$; here we instead disable its 16-dimensional source--estimate interaction $\mathbf c_{m,n}^{\mathrm{src}}$ for target modalities at inference. The other relation components, reconstructed estimates, and observed inputs are retained, allowing us to inspect which target-specific interactions influence the trained fusion process.

\begin{table*}[t]
\centering
\caption{IEMOCAP observed-position similarity to $Z_m$; Mean averages six pairs.}
\label{tab:msrgf_supp_preservation}
\small
\setlength{\tabcolsep}{8pt}
\renewcommand{\arraystretch}{1.08}
\begin{tabular}{@{}lcccc@{}}
\toprule
Setting / target & $\operatorname{MSE}(C_m,Z_m)$ & $\operatorname{MSE}(\widetilde H_m,Z_m)$ & $\operatorname{Cos}(C_m,Z_m)$ & $\operatorname{Cos}(\widetilde H_m,Z_m)$ \\
\midrule
AV / A & 2.033782 & 0.000106 & 0.018657 & 0.999952 \\
AV / V & 2.020127 & 0.000010 & 0.152937 & 0.999995 \\
TA / T & 1.155433 & 0.021737 & 0.031935 & 0.990000 \\
TA / A & 1.148216 & 0.011019 & 0.019081 & 0.995030 \\
TV / T & 2.075091 & 0.022260 & 0.010071 & 0.989525 \\
TV / V & 1.454458 & 0.000786 & 0.159364 & 0.999619 \\
\midrule
Mean & 1.647851 & \textbf{0.009320} & 0.065341 & \textbf{0.995687} \\
\bottomrule
\end{tabular}
\end{table*}

\begin{table}[t]
\centering
\caption{Target-specific source--estimate interaction interventions.}
\label{tab:msrgf_supp_target_relation}
\small
\setlength{\tabcolsep}{6pt}
\renewcommand{\arraystretch}{1.10}
\begin{tabular*}{\columnwidth}{@{\extracolsep{\fill}}lcc@{}}
\toprule
Setting & Target(s) & $\Delta$Acc/F1 \\
\midrule
MOSEI 90\% & All & +0.88/+0.86 \\
IEMOCAP AV & T & +0.35/+0.93 \\
IEMOCAP TV & A & +0.35/+0.39 \\
\bottomrule
\end{tabular*}
\end{table}

The largest displayed IEMOCAP F1 reduction occurs when the interaction for the missing textual target is disabled under AV (0.93 points), while disabling the audio-target interaction under TV reduces F1 by 0.39 points. On MOSEI with 90\% within-modality missingness, removing the interaction for all targets reduces Acc/F1 by 0.88/0.86 points. These interventions localize the effect of the relation term and complement the retrained ablations, which measure the consequences of learning without that information.

\subsection{Difference Refinement}\label{sec:msrgf_supp_refinement}
The reference $C_m$ aggregates the source estimates, whereas the refinement term uses their individually transformed deviations $\Delta_m^n=R_m^n-C_m$. To examine whether the learned update direction matters for prediction, we intervene on $U_m=g_m^{\mathrm{enh}}\odot N_m$ for entirely missing targets in a trained model, keeping the reference and other enhanced modalities unchanged. Table~\ref{tab:msrgf_supp_refinement_task} compares the original update with its zeroed, randomized, reversed, and sample-permuted counterparts; the randomized update preserves its norm, while permutation exchanges updates across samples.

For TV$\rightarrow$A, reversing the update reduces F1 from 80.61 to 79.29, and substituting an equal-norm random direction gives 80.03. A similar ordering is observed for AV$\rightarrow$T, indicating that the task effect depends on the learned update direction in these settings. We then evaluate decoded reconstruction under 90\% within-modality missingness on MOSEI, where the visual target shows the clearest effect: its MSE increases from 0.644883 to 0.651519 without refinement and to 0.666883 when the update is reversed. The acoustic target follows the same ordering, as reported in Table~\ref{tab:msrgf_supp_refinement_mse}.

\begin{table}[t]
\centering
\caption{Decoded target MSE on MOSEI at 90\% within-modality missingness.}
\label{tab:msrgf_supp_refinement_mse}
\small
\setlength{\tabcolsep}{3pt}
\begin{tabular*}{\columnwidth}{@{\extracolsep{\fill}}lcccc@{}}
\toprule
Target & Full & w/o $N_m$ & Random & Reversed \\
\midrule
A & \textbf{1.052734} & 1.054066 & 1.054196 & 1.055929 \\
V & \textbf{0.644883} & 0.651519 & 0.653729 & 0.666883 \\
\bottomrule
\end{tabular*}
\end{table}

To check whether the visual-target reconstruction change is confined to a single missing rate, we compare the full and zero-refinement representations across all nine within-modality rates on MOSEI. Table~\ref{tab:msrgf_supp_rate_geometry} reports a positive decoded-MSE reduction at every rate, averaging 2.103$\times 10^{-3}$ and reaching its largest value at 90\% missingness. This rate-wise result extends the direction intervention from individual test conditions to a broader range of observation coverage.

\begin{table}[t]
\centering
\caption{Decoded visual-target MSE reduction from difference refinement on MOSEI.}
\label{tab:msrgf_supp_rate_geometry}
\setlength{\tabcolsep}{8pt}
\begin{tabular*}{\columnwidth}{@{\extracolsep{\fill}}lc@{}}
\toprule
Within-modality missing rate & MSE reduction ($\times 10^{-3}$) \\
\midrule
10\% & +3.700 \\
20\% & +1.821 \\
30\% & +1.359 \\
40\% & +1.819 \\
50\% & +1.529 \\
60\% & +0.222 \\
70\% & +0.740 \\
80\% & +1.105 \\
90\% & +6.636 \\
\midrule
Avg & \textbf{+2.103} \\
\bottomrule
\end{tabular*}
\end{table}

\subsection{Observed-Feature Preservation}\label{sec:msrgf_supp_observed}
The preceding experiments address missing-target reconstruction, but RGF also needs to retain informative positions that are already observed. To measure this aspect of integration directly, we compare $C_m$ and $\widetilde H_m$ with the projected observations $Z_m$ at positions selected by $M_m$, using the same 128-dimensional representation space. The six IEMOCAP setting--target pairs in Table~\ref{tab:msrgf_supp_preservation} distinguish a reference constructed from source estimates from the final representation after observation-conditioned fusion.

Across these six pairs, the mean MSE to $Z_m$ decreases from 1.647851 for $C_m$ to 0.009320 for $\widetilde H_m$, while mean cosine similarity rises from 0.065341 to 0.995687. These measurements characterize the final integration outcome at observed positions; the retrained \textit{Hard Observation Copy}, \textit{w/o Observed Fusion}, and preservation-loss ablations in Table~\ref{tab:msrgf_ablation_add} separately examine which computational and training choices contribute to task performance. Thus, close agreement with observed features and predictive effectiveness are assessed through complementary measurements.

\subsection{Global Context Modulation}\label{sec:msrgf_supp_gcm_effect}
Finally, we examine the prediction stage after the enhanced modality representations have been constructed. The retrained \textit{w/o GCM} ablation removes global-context construction and modulation during learning; an inference-time intervention instead disables the same contribution in the trained network while retaining RGF outputs and the original predictor. Table~\ref{tab:msrgf_supp_gcm} reports the condition-specific changes, with F1 reductions of 2.457 points under TAV and 0.951 points under AV.

\begin{table}[t]
\centering
\caption{IEMOCAP GCM intervention results.}
\label{tab:msrgf_supp_gcm}
\small
\setlength{\tabcolsep}{5pt}
\begin{tabular*}{\columnwidth}{@{\extracolsep{\fill}}lcc@{}}
\toprule
Setting & Full Acc/F1 & $\Delta$Acc/F1 \\
\midrule
TAV & 81.70/80.81 & +1.887/+2.457 \\
AV & 79.73/77.58 & +0.656/+0.951 \\
\bottomrule
\end{tabular*}
\end{table}

The intervention is also applied to the complete IEMOCAP test set under each missingness protocol, together with the SWR pathway intervention, to distinguish the effect of a combination from the overall protocol average. Table~\ref{tab:msrgf_supp_protocol} shows that disabling SWR has the larger immediate impact across the missingness protocols, while disabling GCM yields smaller but positive changes in both Acc and F1. These results complement the retrained component comparisons and conclude the analysis from source-wise reconstruction through fusion to task prediction.

\begin{table}[t]
\centering
\caption{Inference-time pathway interventions on IEMOCAP.}
\label{tab:msrgf_supp_protocol}
\footnotesize
\setlength{\tabcolsep}{2pt}
\renewcommand{\arraystretch}{1.09}
\begin{tabular*}{\columnwidth}{@{\extracolsep{\fill}}lccc@{}}
\toprule
Disabled path & Fixed & Random whole & Random within \\
\midrule
SWR & +3.71/+3.54 & +4.54/+5.42 & +2.10/+2.43 \\
GCM & +0.41/+0.52 & +0.21/+0.23 & +0.10/+0.11 \\
\bottomrule
\end{tabular*}
\end{table}